\documentclass[%
 reprint,
 amsmath,amssymb,
 aps,
]{revtex4-2}

\usepackage{graphicx}
\usepackage{dcolumn}
\usepackage{bm}

\usepackage{enumitem} 
\usepackage{subfigure}
\usepackage{float}
\usepackage{color}
\usepackage{ulem}
\usepackage{amsmath}
\usepackage{algorithm}
\usepackage{algpseudocode}

\begin{document}


\title{Interfacing branching random walks with Metropolis sampling: \\
constraint release in auxiliary-field quantum Monte Carlo}

\author{Zhi-Yu Xiao}
 \affiliation{%
Department  of  Physics,  College  of  William  \&  Mary,  Williamsburg,  Virginia  23187,  USA
}%

\author{Hao Shi}
\affiliation{
Department of Physics and Astronomy, University of Delaware, Delaware, 19716, USA
}%
\author{Shiwei Zhang}
\affiliation{%
Center  for  Computational  Quantum  Physics,  Flatiron  Institute,  New  York,  NY  10010,  USA
}%
\date{\today}

\begin{abstract}
We present an approach to interface branching random walks with Markov chain Monte Carlo sampling, and to switch seamlessly between the two. The approach is discussed in the context of auxiliary-field quantum Monte Carlo (AFQMC) but is applicable to other Monte Carlo calculations or simulations. In AFQMC, the formulation of branching random walks along imaginary-time is needed to realize a constraint to control the sign or phase problem. The constraint is derived from an exact gauge condition, and is in practice implemented approximately with a trial wave function or trial density matrix, which can break exactness in the algorithm. We  use the generalized Metropolis algorithm to sample a selected portion of the imaginary-time path after it has been produced by the branching random walk. This interfacing allows a constraint release to follow seamlessly from the constrained-path sampling, which can reduce the systematic error from the latter. It also provides a way to improve the computation of correlation functions and observables that do not commute with the Hamiltonian. We illustrate the method in atoms and molecules, where improvements in accuracy can be clearly quantified and near-exact results are obtained. We also discuss the computation of the variance of the Hamiltonian and propose a convenient way to evaluate it stochastically without changing the scaling of AFQMC. 

\end{abstract}

\pacs{Valid PACS appear here}

\maketitle

\section{\label{sec:level1}INTRODUCTION}

The study of interacting quantum many-body systems represents one of the main challenges in areas including condensed matter physics, nuclear physics, cold atoms physics, as well as quantum chemistry and materials science. Quantum many-body systems are characterized by their high degree of complexity and the interplay between the various degrees of freedom. A general approach does not yet exist which can treat the full complexity of an interacting quantum system and yields systematically accurate results across different ranges of many-body models and materials. As a result, different methods are used with specialization to focus on different systems or different aspects of the same system. Each class of methods still has limitations, but progress in development has been accelerating, in significant part due to recent efforts in benchmark and collaborative studies \cite{simons_material_2020, Hubbard_benchmark_2015}. The continued development of more general and more accurate numerical methodologies is instrumental in meeting the challenges of understanding and predicting the properties of interacting quantum systems.

Quantum Monte Carlo approaches \cite{becca_QMC_correlated_2017, ceperley_chemistry_1995, foulkes_solids_2001, RevModPhys_nuclear_2015} represent one important class of many-body methods, which have been applied in a number of areas in the study of correlated quantum systems. Equilibrium state QMC samples the many-body wave function/density matrix in a chosen basis space. Depending on the form of the Hamiltonian, the goal of the computation (e.g., finite-temperature versus ground state), the choice of the basis space, etc., QMC has many flavors, and the Monte Carlo sampling takes different forms. The vast majority of applications in non-trivial quantum systems require elaborate specialization of the Monte Carlo technique in order to achieve sufficient sampling efficiency. 

There are two primary algorithmic structures to realize the Monte Carlo sampling: Markov chain Monte Carlo (MCMC) or branching random walks (BRW). One chooses one algorithm or the other, and the entire framework of the QMC calculation then stays within that choice. The difference between the choices is sometimes only in efficiency or convenience, while other times it is more fundamental. For example, with auxiliary-field quantum Monte Carlo (AFQMC), it is convenient to use MCMC under a path-integral formalism when there is no sign problem or when one can have an acceptable average sign or phase to obtain useful results. This approach is seen widely in lattice QCD \cite{AFQMC_L_QCD}, and in condensed matter \cite{QMC_DQMC} (under what is often referred to as determinantal quantum Monte Carlo, DQMC). On the other hand, when the sign/phase problem is more severe, a class of algorithms to control them by imposing sign or gauge conditions has proved highly effective and reliable \cite{simons_material_2020, Hubbard_benchmark_2015}, which are seeing growing applications in condensed matter \cite{Mario_hydrogen_chain,Bo_Xiao_Stripe_Hubbard} and quantum chemistry \cite{Joonho_chemistry}. Because of ergodicity problems with MCMC, the imposition of the constraint requires breaking up the construction of the imaginary-time paths with the use of BRW along imaginary time.

In the present work, we address the issue of how to interface the two sampling methods. In particular, we show how samples from BRW can be used to seamlessly initialize a MCMC sampling. Although technical, this has important conceptual and practical implications. One significant area of application is that this approach allows, following a constrained path or phaseless AFQMC calculation, constraint release, which can systematically reduce the residual constraint bias without requiring the change of a trial wave function or any other external input.

Constraint release has been discussed in previous works \cite{Hao_symmetry, Ankit_AFQMC_freeRelease}, in which one continues a release calculation with BRW, following a constrained calculation under the same framework. The basic idea is straightforward and is similar to the schemes used in diffusion Monte Carlo for performing released node calculations following a fixed-node calculation \cite{RN-DMC}. However, the interface between the constrained and unconstrained calculations involves a change in importance sampling strategies, which is, in practice, unstable. Of course, as one relaxes the constraint, the instability from the sign or phase problem returns, and the calculations become intrinsically less stable (growing sign or phase) as a function of imaginary time for the release. Here we refer to the {\it additional\/} numerical instability at the interface between the constrained and unconstrained portions, where one must change the importance function and introduce large noise. The method introduced here removes this problem. We combine BRW with MCMC sampling and adopt it in the AFQMC framework as a constraint-releasing procedure. By constructing a portion of the imaginary-time path with BRW and then updating them with Metropolis sampling, we can maintain the importance sampling structure inherited from BRW.

An alternative view of the algorithm we introduce is that it allows the standard ``exact'' QMC in which one keeps track of the sign to incorporate the best initial state. For ground-state calculations, this means we perform a free projection calculation under MCMC sampling, using an initial state produced by a constrained calculation (e.g., constrained path \cite{QMC_Zhang_Constrained_1997} or phaseless AFQMC \cite{AFQMC_Zhang-Krakauer-2003-PRL}). For finite-temperature calculations, the algorithm allows the MCMC sampling to initialize from a highly accurate but approximate many-body density matrix obtained from constrained path AFQMC \cite{zhang1999finite, Yuan-Yao_FT_AFQMC}.

The idea of interfacing BRW with MCMC, for example, using the former as a starting point in the latter, can also be useful in other contexts. Branching random walk algorithms, similar to ray tracing \cite{keller_RayTracing} or neutron transport simulations \cite{neutron_transport_QMC}, typically require importance sampling to be efficient in high dimensions. Compared to MCMC sampling, they are typically less prone to ergodicity problems. Thus they could serve as an effective pre-sampler for MCMC, provided one could undo the importance sampling seamlessly. The method discussed in this paper concerns the optimal way to undo the importance function, and can be useful for these applications as well.

The rest of this paper is organized as follows. In Sec.~\ref{sec:level2}, we provide an overview of the formalism of AFQMC in such a way that facilitates the discussion of the connection and difference between the MCMC and BRW algorithms. In Sec.~\ref{sec:level3}, we introduce the formalism of Metropolis release constraint in ground-state calculations, and discuss extensions and other applications of the algorithm, including in computing the variance of the Hamiltonian and finite-temperature calculations. In Sec.~\ref{sec:level4}, we present illustrative results of Metropolis AFQMC in atoms/molecules. In Sec.~\ref{sec:level5}, we conclude and propose potential future works.

\section{\label{sec:level2} Preliminaries}

\subsection{\label{sec:level21} Overview}

Our discussions start from the Hamiltonian in ``Monte Carlo form" \cite{Hao_Some_recent_developments}:
\begin{equation}
\hat{H}= \hat{T}+\frac{1}{2}\sum^\Gamma_\gamma \hat{L}_\gamma ^2 
\label{eq:H_MC}
\end{equation}
where $\hat{T}$ and $\hat{L}_\gamma$ denote sets of one-body operators whose matrix elements are explicitly specified in the chosen basis. 

In this paper, we will focus on ground-state calculations, but we comment on finite temperature in Sec.~\ref{sec:level3}. One way to reach the ground state of generic Hamiltonian is to proceed with imaginary-time projection:
\begin{equation}
|\Psi_{0}\rangle\propto \underset{\beta \to \infty}{\lim} e^{-\beta\hat{H}}|\Psi_I\rangle\,,
\end{equation}
which propagates the given initial state $|\Psi_I\rangle$ to the ground state $|\Psi_{0}\rangle$ of the Hamiltonian $\hat{H}$,
if the overlap $\langle\Psi_{0}|\Psi_I\rangle$ is non-zero. 
For the ensuing discussions, it is convenient to express the computational implementation in the following form:
\begin{equation}
\langle \hat{A}\rangle = \frac{\langle \Psi_T| e^{-\beta_L \hat{H}} \hat{A} e^{-\beta_R \hat{H}} |\Psi_{I}\rangle}{\langle \Psi_T|e^{-\beta_L \hat{H}}\,e^{-\beta_R \hat{H}}|\Psi_{I}\rangle}\,,
\label{eq:overview1}
\end{equation}
where $\beta_L+\beta_R=\beta$, and $\hat{A}$ is an observable or equal-time correlation function whose expectation value with respect to the ground state is being computed. (Imaginary-time-dependent correlations can also be computed, but for the purpose of illustrating the different algorithms, we will restrict ourselves to the simpler form.) 

Different flavors of ground-state quantum Monte Carlo (QMC) can be thought of as different ways of evaluating the right-hand side of Eq.~(\ref{eq:overview1}). We have deliberately used two different states $|\Psi_T\rangle$ (for ``trial'') and $|\Psi_I\rangle$ (for ``initial'') to denote the left- and right-hand sides. The simplest default choice is to use the same, for example, a single Stater determinant from mean-field or density-functional calculations, but the two sides can have different characteristics because of the actual implementation of the Monte Carlo algorithm. A better choice which is a better approximation to the true ground state, will allow smaller choices of $\beta_L$ and $\beta_R$, which is crucial in the presence of a sign or phase problem, as we further discuss below.

Auxiliary-field QMC (AFQMC) chooses to perform the imaginary-time projection as path-integrals in auxiliary-field space. To realize this, one uses the Suzuki-Trotter decomposition \cite{Suzuki,Trotter} to break the imaginary-time evolution operator into time slices 
\begin{equation}
e^{- \beta_L\hat{H} }  =   (e^{-\tau \hat{H} })^m\,, 
\qquad 
e^{- \beta_R\hat{H} }  =   (e^{-\tau \hat{H} })^n\,,
\end{equation}
where $m= \beta_L/\tau $ and  $n= \beta_R/\tau $ define the necessary number of time slices given a choice of the time step $\tau$. As seen below, the time step must be sufficiently small to make the errors from commutators negligible (compared to MC statistical errors); in practice, an extrapolation is often performed with separate calculations using several choices of $\tau$. One then applies the Hubbard-Stratonovich transformation \cite{HS_transformation_H, HS_transformation_S} to decouple the short-time propagator in each slice
\begin{equation}
e^{-\tau \hat{H} }\approx  \int\mathrm{d}\textbf{x}\, p(\textbf{x})\,\hat{B}(\textbf{x})\,,
\label{eq:HS}
\end{equation}
where $p(\textbf{x})$ is a probability density function, either discrete or continuous such as
\begin{equation}
p(\textbf{x})= \prod_\gamma \frac{1}{\sqrt{2\pi}}e^{-x_\gamma^2/2}\propto e^{-\textbf{x}\cdot\textbf{x}/2},
\end{equation}
and the one-body propagator 
\begin{equation}
\hat{B}(\textbf{x})=e^{-\tau\hat{T}/2}e^{\sum_\gamma x_\gamma\sqrt{-\tau}\hat{L}_\gamma }e^{-\tau\hat{T}/2}.
\label{eq:HS-B}
\end{equation}
where $\textbf{x}$ indicates a series of $\{x_\gamma\}$ with $x_\gamma \in \mathbb{R}$ and $\textbf{x}\cdot\textbf{x}=\sum_{\gamma}x_\gamma^2$. 

\begin{figure}[htbp]
\includegraphics[scale = 0.28]{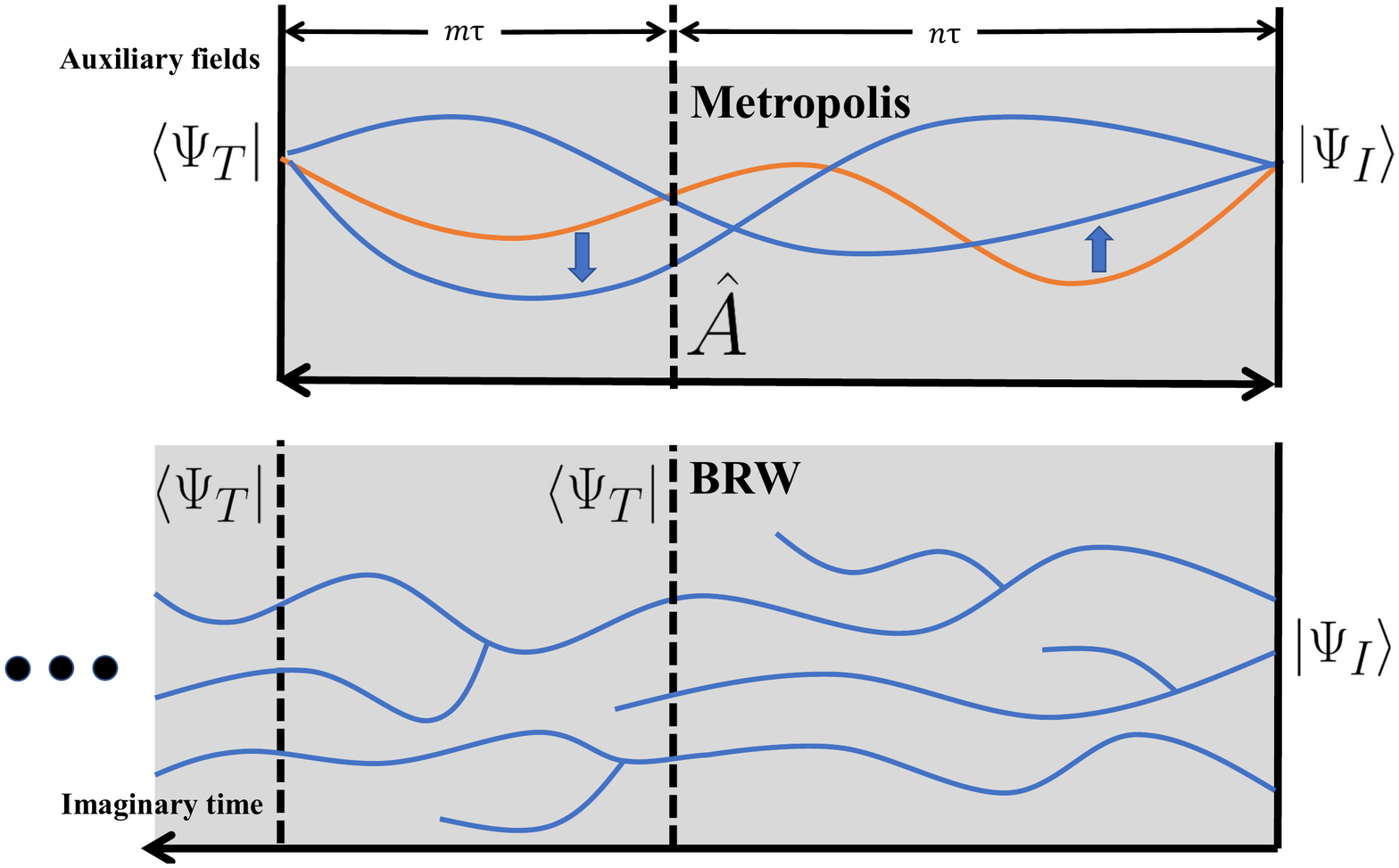} 
\caption{Illustrations of the Metropolis  (top) and branching random walk (bottom) algorithms in AFQMC. The solid curves represent the sampled paths ${\textbf{X}}$ that connect $\langle \Psi_I |$ to $|\Psi_T\rangle$. In the Metropolis algorithm, the auxiliary fields are sampled by updating from an existing path. Observables $\hat{A}$ are measured in the middle of the chain. In the branching random walk algorithm, a population of walkers are propagated from right to left along the direction of imaginary time. The random walks are open-ended in ground-state methods. 
}
\label{Fig.BRW_MCMC} 
\end{figure}

We now return to Eq.~(\ref{eq:overview1}), and rewrite it explicitly as:
\begin{widetext}
\begin{equation}
\langle \hat{A}\rangle 
=\frac{\langle \Psi_T| e^{-m\tau \hat{H}} \hat{A} e^{-n\tau \hat{H}} |\Psi_{I}\rangle}{\langle \Psi_T|e^{-(m+n)\tau \hat{H}}|\Psi_{I}\rangle}
=\frac{\int \prod_{i=1}^{m+n}\,d\textbf{x}^{i}\,p(\textbf{x}^{i})\langle \Psi_T|\prod_{i=n+1}^{m+n}B(\textbf{x}^{i})\,\hat{A}\,\prod_{i=1}^{n} B(\textbf{x}^{i})|\Psi_{I}\rangle}{\int \prod_{i=1}^{m+n}\,d\textbf{x}^{i}\,p(\textbf{x}^{i})\langle \Psi_T|\prod_{i=1}^{m+n}B(\textbf{x}^{i})|\Psi_{I}\rangle }\, ,
\label{eq:overview2}
\end{equation}
\end{widetext}
which we can think of as a total projection time with length $\beta= \beta_L+\beta_R$, and the different configurations of $m$ and $n$, with  $(m+n)$ fixed, as different locations to perform the measurements. For the energy, the different locations should all lead to the same results in their statistical mean. For operators $\hat{A}$ which do not commute with the Hamiltonian, the answer is only converged when $\beta_L$ and $\beta_R$ reach a threshold ``equilibration time'' which depends on $|\Psi_T\rangle$ and $|\Psi_I\rangle$, respectively. 
There are basically two flavors of ground-state AFQMC. From a high-level algorithmic standpoint, they can be summarized as
\begin{itemize}
    \item Metropolis or Markov Chain Monte Carlo sampling, as illustrated in Fig.~\ref{Fig.BRW_MCMC} (top). In this approach, $\beta$ is pre-chosen and fixed throughout the simulation (i.e., $(m+n)$ is fixed), and the MC sampling carries an object $X=\{\textbf{x}^{m+n}, \cdots,\textbf{x}^{1}\}$, proposes updates on this path $X\rightarrow X'$, and accepts/rejects the proposal. Typically one sweeps through the path (right-to-left or vice versa) to make proposals. The algorithm takes $|\Psi_T\rangle$ and $|\Psi_I\rangle$ as single Slater determinants. More complicated wave functions can be used via Monte Carlo sampling.  
    \item BRW, as illustrated in Fig.~\ref{Fig.BRW_MCMC} (bottom). In this approach, $\beta$ is not fixed, and sampling is conducted by open-ended random walks.
    The MC sampling typically carries a population indexed by $k$, and the branching random walk proceeds from right to left: $\{\textbf{x}_k^{1}\}\rightarrow \{\textbf{x}_k^{2}\} \rightarrow \cdots $.
\end{itemize}

In the next subsection, we outline the algorithm for importance-sampled BRW in more detail in a way that will facilitate the ensuing discussion on how to interface it with the MCMC.

\subsection{\label{sec:level22} AFQMC: BRW, importance sampling, and constraint}

In the Metropolis path-integral formalism, it is difficult to impose a constraint without incurring a global ergodicity problem. In order to deal with the sign and phase problem, it was necessary to reformulate AFQMC in the form of BRW \cite{QMC_Zhang_Constrained_1997, AFQMC_Zhang-Krakauer-2003-PRL,lecturenotes-2019}. The most straightforward way to think about the branching random walk is free projection. In this framework, with Eq.~(\ref{eq:HS}), the ground-state wavefunction throughout the imaginary-time projection can be represented as
\begin{equation}
\begin{aligned}
e^{- n\tau \hat{H} }|\Psi_I\rangle 
\rightarrow 
|\Psi^n\rangle = 
\int \mathrm{d}\textbf{X} p(\textbf{X}) |\phi^n(\textbf{X})\rangle\,,
\end{aligned}
\label{eq:BRW}
\end{equation}
where $\textbf{X} = (\textbf{x}^n \cdots  \textbf{x}^i \cdots  \textbf{x}^1)$ is a path of auxiliary fields and $|\phi^n(\textbf{X})\rangle= \prod_{i=1}^{n} B(\textbf{x}^{i})|\Psi_I\rangle$ is a single Slater determinant. We will use uppercase symbols to denote many-body wave functions, such as $|\Psi^n\rangle$ vs.~$|\phi^n\rangle$. (As mentioned, $|\Psi_I\rangle$ and  $|\Psi_T\rangle$ can both be many-body wave functions, for example, a linear combination of Slater determinants --- hence their symbol choice, even though we assumed they are of single determinant form in the illustrations.)

Equation~.\ref{eq:BRW} maps the imaginary-time projection to random walks in the manifold of Slater determinants along the direction of imaginary time evolution. At each time step $i$, a new configuration $\textbf{x}^i$ is sampled with a probability distribution given by $p(\textbf{x}^i)$. The orbitals in the Slater determinants cease being orthonormal during the propagation, which can be thought of as different walkers contributing different ``weights'' to $|\Psi^n\rangle$ in Eq.~(\ref{eq:BRW}).

The free-projection is inefficient as a sampling method since the different paths leading to different $|\phi^n\rangle$'s are sampled randomly and uniformly. Furthermore, it is incompatible with the imposition of a constraint, especially when the Hubbard-Stratonovich transformation in Eq.~(\ref{eq:HS}) leads to complex phases. This second point is nuanced and more fundamental and involves separating out the (legitimate) contribution of the complex phases from the two-body interaction from the random phase fluctuations, which are the origin of the phase problem \cite{AFQMC_Zhang-Krakauer-2003-PRL,lecturenotes-2019}. These difficulties are dealt with by introducing an importance sampling transformation. 

Under importance sampling, the many-body wave function, as it is propagated in imaginary time, can be thought of as:
\begin{equation}
 |\Psi^i\rangle
\propto \sum_k W^i_k\,\frac{|\phi^i_k \rangle}{\langle \Psi_T|\phi^i_k \rangle}\,, 
\label{eq:WF-imp}
\end{equation}
where $\{|\phi_k^i\rangle, W_k^i\}$ denote the 
population of random walkers, labeled by $k$, at imaginary time step $i$. 
This can be realized by a dynamic shift in the integration (or sum) over auxiliary fields in Eq.~(\ref{eq:HS}):
\begin{equation}
e^{-\tau \hat{H} }\approx  \int\mathrm{d}\textbf{x}\, p(\textbf{x}- \overline{\textbf{x}})\,\hat{B}(\textbf{x}- \overline{\textbf{x}})\,,
\label{eq:HS-shifted}
\end{equation}
with which each step of the imaginary-time projection $e^{-\tau \hat{H} }|\Psi^i\rangle \rightarrow |\Psi^{i+1}\rangle$ can be performed while preserving the 
structure of the wave function in Eq.~(\ref{eq:WF-imp}):
\begin{equation}
e^{-\tau \hat{H} }\,
\sum_k W^{i}_k \frac{|\phi^{i}_k \rangle}{\langle \Psi_T|\phi^{i}_k \rangle} 
 \rightarrow 
 \sum_k W^{i+1}_k \frac{|\phi^{i+1}_k \rangle}{\langle \Psi_T|\phi^{i+1}_k \rangle}\,.
\label{eq:prop-imp}
\end{equation}
Formally this is given by the following propagation of each $|\phi^{i}_k \rangle$:
\begin{equation}
 \int\mathrm{d}\textbf{x}\, p(\textbf{x})\,I(\textbf{x}, \overline{\textbf{x}}, \phi)\,\hat{B}(\textbf{x}- \overline{\textbf{x}}) |\phi\rangle \,,
\label{eq:full-prop-imp-samp}
\end{equation}
where the importance function is defined as
\begin{equation}
\begin{aligned}
I(\textbf{x}, \overline{\textbf{x}}, \phi) = \frac{p(\textbf{x}-\overline{\textbf{x}})}{p(\textbf{x})}\frac{\langle \Psi_T| \hat{B}(\textbf{x} - \overline{\textbf{x}})| \phi \rangle }{\langle \Psi_T|\phi \rangle }\,.
\end{aligned}
\label{eq:importance_function}
\end{equation}
As mentioned, the shift $\overline{\textbf{x}}$ is dynamic and depends on the current position of the walker, $|\phi\rangle$.
It is given component by component (see Eq.~(\ref{eq:HS-B})), chosen to minimize the fluctuation of the weights. Often referred to as a force bias, the optimal shift 
for small $\tau$ is given by \cite{AFQMC_shift}:
\begin{equation}
\begin{aligned}
\overline{\textbf{x}}_\gamma =-\sqrt{\tau}\frac{\langle  \Psi_T|\hat{L}_\gamma |\phi \rangle}{\langle \Psi_T| \phi\rangle}\,.
\end{aligned}
\label{eq:shift1}
\end{equation}
In Eqs.~\ref{eq:full-prop-imp-samp}, \ref{eq:importance_function} and \ref{eq:shift1}
we have omitted the superscript $i$ and subscript $k$ in $\phi$ and $\overline{\textbf{x}}$ to reduce clutter.
The random walk step involves
\begin{enumerate}
    \item sample a field from $p(\textbf{x})$\,,
    \item advance the walker $\hat{B}(\textbf{x} - \overline{\textbf{x}}_k^i)| \phi_k^i \rangle
    \rightarrow | \phi_k^{i+1} \rangle$\,,
    \item assign weight $I(\textbf{x}, \overline{\textbf{x}}_k^i, \phi_k^i)\,W_k^i \rightarrow W_k^{i+1}$\,.
\end{enumerate}

The ground-state energy can be computed using the mixed estimator \cite{AFQMC_Zhang-Krakauer-2003-PRL,WIREs_Mario},
i.e., setting $m=0$ and measuring at any $n$ for which $\beta_R>\overline{\beta}_{I\rightarrow {\rm ph}}$, where $\overline{\beta}_{I\rightarrow {\rm ph}}$ is the projection time needed to reach the phaseless ground state $|\Psi_0^{\rm ph}\rangle$ from $|\Psi_I\rangle$. To realize Eq.~(\ref{eq:overview2}) for a general observable, an additional back-propagation scheme \cite{Wirawan_boson,AFQMC_PathRestoration} can be used, in which the random walk generates $|\Psi^n\rangle$ after $n$ steps, with 
\begin{equation}
|\phi_k^n\rangle = \prod_{i=1}^{n} 
\hat{B}(\textbf{x}_k^i - \overline{\textbf{x}}_k^i)\,|\Psi_I\rangle\,,
\label{eq:BP-fwd}
\end{equation}
where with the $\prod$ we imply a time order from right to left when operators are involved. The random walk then continues for $m$ steps, producing a path of auxiliary fields which lead to final weights $W_k^{m+n}$:
\begin{equation}
W_k^{m+n} = \prod_{i=1}^{m+n} I(\textbf{x}^i, \overline{\textbf{x}}^i, \phi_k^i)
= \prod_{i=n+1}^{m+n} I(\textbf{x}^i, \overline{\textbf{x}}^i, \phi_k^i)\, W_k^{n}\,.
\label{eq:BP-wt}
\end{equation}
Then we back-propagate from $\langle \Psi_T|$ along the sampled paths to generate
\begin{equation}
\langle \tilde \phi_k^m| = \langle \Psi_T|\,\prod_{i=1}^{m} 
\hat{B}(\textbf{x}_k^{n+i} - \overline{\textbf{x}}_k^{n+i})\,,
\label{eq:BP-bra}
\end{equation}
with which observable can be computed as
\begin{equation}
\langle \hat{A} \rangle \approx \frac{\sum_k W^{n+m}_{k} \frac{\langle \tilde \phi_k^m|\hat{A}|\phi^{n}_k \rangle}{\langle \tilde \phi_k^m|\phi^{n}_k \rangle}}{\sum_k W^{n+m}_{k}}\,.
\label{eq:BP-obs}
\end{equation}
Note that the mixed estimator is a special case of Eq.~(\ref{eq:BP-obs}) with $m=0$.

The above branching random walk formalism is, in principle, equivalent to the Metropolis path integral approach. Absent of a sign of phase in the importance function in Eq.~(\ref{eq:importance_function}) (i.e., if $I(\textbf{x}, \overline{\textbf{x}}, \phi) >0$ for all paths), the branching random walk algorithm is exact and is simply an alternative to the latter, with their relative efficiency depending on the landscape of the field space and the circumstances of the problem. 

When there is a sign or phase problem, both methods will encounter difficulties. The branching random walk with importance sampling is formulated to make it natural and convenient to impose a constraint. For example, the phaseless AFQMC (ph-AFQMC) method imposes a gauge condition on the walkers \cite{AFQMC_Zhang-Krakauer-2003-PRL,lecturenotes-2019}. This can be viewed as modifying the importance function:
\begin{equation}
I_{\textup{ph}}(\textbf{x}, \overline{\textbf{x}}, \Psi) = \bigl|I(\textbf{x}, \overline{\textbf{x}}, \Psi)\bigr|\,\cdot \textup{max}(0, \textup{cos}(\Delta \theta))\\
\label{eq:importance_function_CP}
\end{equation}
with 
\begin{equation}
\Delta\theta=\textup{Arg}\frac{\langle \Psi_T| \hat{B}(\textbf{x} - \overline{\textbf{x}})| \Psi \rangle }{\langle \Psi_T|\Psi \rangle }.
\end{equation}
Implementation details can vary (e.g., local energy vs.~hybrid \cite{hybrid_phaseless}, removing the abs by retaining the overall phase for long projection time \cite{AFQMC_Zhang-Krakauer-2003-PRL,AFQMC_PathRestoration}, cos projection vs.~half plane or other ways to eliminate the finite density at the origin in the complex plane \cite{Hao_Some_recent_developments}). When the auxiliary fields are all real (e.g., for Hubbard interactions), the phaseless formalism reduces to the constrained path approach \cite{QMC_Zhang_Constrained_1997}, where $\Delta \theta=0$ and the $0$ values in the importance function keeps the random walk free of sign problem. 
The modification of importance function $I_{\textup{ph}}(\textbf{x}, \overline{\textbf{x}}, \Psi)$ introduce a systematic bias which depends on $\langle \Psi_T|$. A large body of benchmark studies has shown that AFQMC with constraint is highly accurate, both in model systems and in real materials \cite{simons_material_2020, Hubbard_benchmark_2015}.

\section{\label{sec:level3} Constraint Release with Metropolis Sampling}

\subsection{\label{sec:level31} Formalism}

In this section, we introduce our approach to interface the BRW seamlessly with the Metropolis algorithm. One way to view this approach is that it allows a Metropolis Monte Carlo calculation of Eq.~(\ref{eq:overview2}) using as ``initial state'' $|\Psi_I\rangle$ a sampled wave function,  in the form of Eq.~(\ref{eq:WF-imp}), which has been produced from the constrained path or phaseless AFQMC:
\begin{equation}
\langle \hat{A}\rangle 
=\frac{\langle \Psi_T| e^{-m\tau \hat{H}} \hat{A} e^{-n\tau \hat{H}} |\Psi^{i}\rangle}{\langle \Psi_T|e^{-(m+n)\tau \hat{H}}|\Psi^{i}\rangle}\,,
\label{eq:MCR-me}
\end{equation}
Because  
$|\Psi^{i}\rangle$ gives an excellent approximation to the true $|\Psi_0\rangle$, this gives an optimal initial state, which minimizes the sign or phase problem as it minimizes $\beta_R$. Performing such calculations with increasing $\beta$ is thus a way to perform a constraint release. As we show below, compared to the more standard approach to perform constraint release \cite{Hao_symmetry, sorella_phase}, this method does not require any modification of the importance function or undoing of the importance sampling, and thus overcomes a fundamental instability. 

In Sec.~\ref{sec:level21} and particularly Sec.~\ref{sec:level22}, we have formulated the branching random walk with importance sampling in such a way as to connect it naturally to the Metropolis framework, which can now be presented in a straightforward manner. Although we use the AFQMC formalism for concreteness, the essential idea is applicable to any such situation, namely how to import a population from an importance-sampled branching random walk into the MCMC framework and continue the sampling. 

\begin{figure}[htbp]
\includegraphics[scale = 0.28]{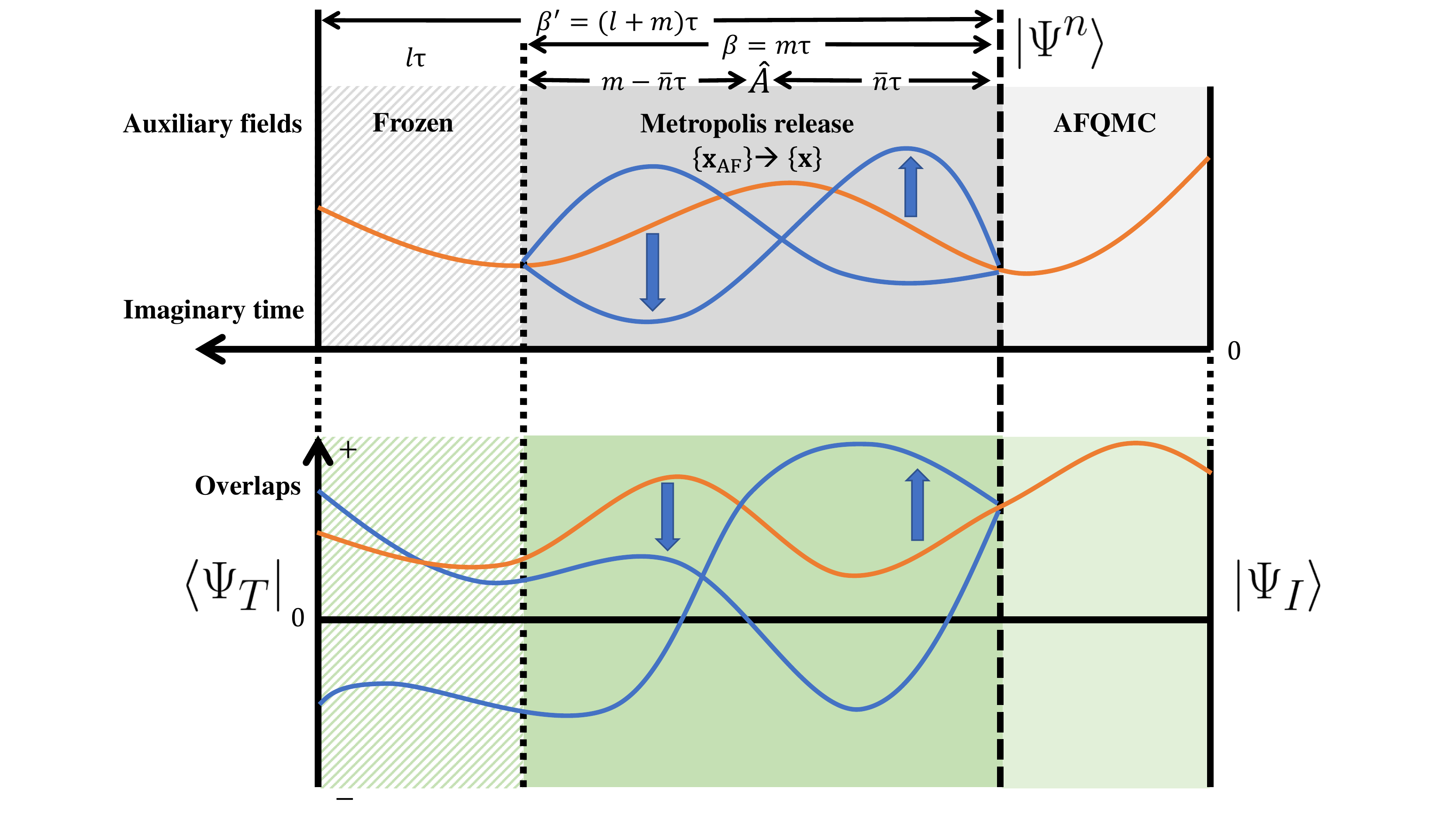} 
\caption{Illustration of the Metropolis release constraint. The top panel shows the paths in auxiliary-field space, ${\mathbf x}^i$, versus $i$ while the bottom panel shows the corresponding overlap $\langle \Psi_T|\prod^{l+m+n}_{i=1} B({\mathbf x}^i) |\Psi_I\rangle$. The solid orange curve represents one path sampled by phaseless (or constrained path) AFQMC. Only the middle segment, the region of $\beta=m\tau$,  is sampled in the Metropolis release. As paths are updated and move away from the original ph-AFQMC configuration (${\mathbf x}_{\rm AF})$, the corresponding overlap, which was originally positive from ph-AFQMC, can turn negative/complex, to allow constraint release.
}
\label{Fig.AFQMC_Metropolis} 
\end{figure}

For notational compactness, let us take $n=0$ in Eq.~(\ref{eq:MCR-me}). There is no loss of generality since, as discussed below, we can perform the measurements of $\hat A$ at different imaginary-time locations along the path generated by the MCMC. For computing the energy, the different locations are equivalent so that the required $m$ for convergence is given by the equilibration time from the phaseless or constrained path ground state to $|\Psi_0\rangle$, i.e., $\beta_L> \overline{\beta}_{{\rm ph}\rightarrow0}$. For computing a general observable, full convergence requires an $m$ set by the total of the above plus the equilibration time needed to converge $|\Psi_T\rangle$ to $|\Psi_0\rangle$:  $\beta_L> \overline{\beta}_{{\rm ph}\rightarrow0}+\overline{\beta}_{{\rm T}\rightarrow0}$, which can be substantially larger since  $\overline{\beta}_{{\rm T}\rightarrow0} > \overline{\beta}_{{\rm ph}\rightarrow0}$ for likely any trial wave function. We now switch the index ``$i$'' in Eq.~(\ref{eq:MCR-me}) to ``$n$'' to be more concrete since $i$ was used as a running index for the ph-AFQMC steps. This suggests yet another way to think about the approach we are discussing, namely to replace $e^{-n\tau \hat H}\,|\Psi_I\rangle$ in Eq.~(\ref{eq:overview2}) with $|\Psi^n\rangle$ from Eq.~(\ref{eq:WF-imp}). Equation~(\ref{eq:MCR-me}) can now be written as
\begin{equation}
\begin{aligned}
\langle \hat{A} \rangle =\frac{\sum_k W^{n}_k\frac{\langle \Psi_T|e^{-m\tau \hat{H}}\hat{A}|\phi^{n}_k \rangle}{\langle \Psi_T|\phi^{n}_k\rangle}}{\sum_k W^{n}_k\frac{\langle \Psi_T|e^{-m\tau \hat{H}}|\phi^{n}_k\rangle}{\langle \Psi_T|\phi^{n}_k\rangle}}\,,
\end{aligned}
\label{eq:overview3}
\end{equation}
where, again, the superscript $n$ indicates that the kets are produced after phaseless projection with a large $n$. 

We now consider the denominator in Eq.~(\ref{eq:overview3}). Returning to Eq.~(\ref{eq:importance_function}) and applying Eq.~(\ref{eq:HS-shifted}), we see that, for each 
$\phi_k^n$
\begin{equation}
 \int\mathrm{d}\textbf{x}\, p(\textbf{x})\,I(\textbf{x}, \overline{\textbf{x}}_k^n, \phi_k^n)= \frac{\langle \Psi_T|e^{-\tau \hat{H}}|\phi^{n}_k\rangle}{\langle \Psi_T|\phi^{n}_k\rangle}\,.
\label{eq:future-wt-step1}
\end{equation}
In the branching random walk, for each $\phi_k^n$, we can imagine sampling many new auxiliary-field configurations, $\textbf{x}$, each of which will lead to a new weight (and a new $\phi_k^{n+1}(\textbf{x})$), as outlined in step \#3 in Sec.~\ref{sec:level22}. The total weights of all the new samples (divided by $W_k^n$) provide a Monte Carlo evaluation of the left-hand side. Applying this recursively, as in Eq.~(\ref{eq:BP-wt}), we have 
\begin{equation}
\frac{W^{n+m}_k}{W^{n}_k} = \frac{\langle \Psi_T|e^{-m\tau \hat{H}}|\phi^{n}_k\rangle}{\langle \Psi_T|\phi^{n}_k\rangle}\,.
\label{eq:future-wt-step-m}
\end{equation}

In the branching random walk, a population of walkers proceed at the same time. The total weight  of all the descendants of each walker measures its relative contribution in the path integral. This is exactly the idea of back-propagation as given in Eq.~(\ref{eq:BP-obs}).

Equation~(\ref{eq:future-wt-step-m}) allows us to rewrite Eq.~(\ref{eq:overview3}) in a new form
\begin{equation}
\langle \hat{A} \rangle \approx \frac{\sum_k W^{m+n}_k\frac{\langle \Psi_T|e^{-m\tau \hat{H}}\hat{A}|\phi^{n}_k \rangle}{\langle \Psi_T|e^{-m\tau \hat{H}}|\phi^{n}_k\rangle}}{\sum_k W^{m+n}_k}
\,.
\label{eq:Metrop-release}
\end{equation}
In other words, the computation of $\langle \hat A\rangle$ is turned into a weighted sum over a collection of 
\begin{equation}
\langle \hat{A} \rangle_k \equiv\frac{\langle \Psi_T|e^{-m\tau \hat{H}}\hat{A}|\phi^{n}_k \rangle}{\langle \Psi_T|e^{-m\tau \hat{H}}|\phi^{n}_k\rangle}
\,.
\label{eq:Metrop-release-est}
\end{equation}
This is just a special realization of Eq.~(\ref{eq:overview2}), with $|\Psi_I\rangle$ chosen as one walker from the phaseless importance sampled ground-state wave function, and $n=0$. Indeed for a general observable $\hat A$, as we discuss in Sec.~\ref{sec:observable}, we will choose to move the measurement location along the path, which eliminates the $n=0$ specialization, so that Eq.~(\ref{eq:Metrop-release-est}) becomes the same as Eq.~(\ref{eq:overview2}), with a particular choice of $|\Psi_I\rangle=|\phi_k^n\rangle$. Thus each $\langle \hat{A} \rangle_k$ can be computed straightforwardly by separate and independent path-integral calculations, using MCMC, for example. 

The MCMC computation of Eq.~(\ref{eq:Metrop-release-est}) for each $|\phi_k^n\rangle$ is independent (of other $k$ values). As usual we turn the propagations of $e^{-\tau \hat H}$ into integrals over auxiliary fields using either Eq.~(\ref{eq:HS}) or Eq.~(\ref{eq:HS-shifted}):
\begin{equation}
\langle \hat{A}\rangle_{k}= \frac{\int d{\mathbf X} 
A_k({\mathbf X})\,\Big|D_k({\mathbf X})\Big|\, s_k({\mathbf X}) }{\int 
d{\mathbf X}\,\Big|D_k({\mathbf X})\Big|\,s_k({\mathbf X}) }\,,
\label{eq:MCMC_A_k}
\end{equation}
where
\begin{equation}
D_k({\mathbf X})= \prod_{i=1}^{m}\,p(\textbf{x}^{i})\langle \Psi_T| 
 \prod_{i=1}^{m}B(\textbf{x}^{i})|\phi^{n}_k\rangle,
\end{equation}
and the ``local'' observable is defined as 
\begin{equation}
A_k({\mathbf X})= \langle \Psi_T| 
 \prod_{i=1}^{m}B(\textbf{x}^{i})\,\hat A |\phi^{n}_k\rangle/\langle \Psi_T| 
 \prod_{i=1}^{m}B(\textbf{x}^{i}) |\phi^{n}_k\rangle,
\end{equation}
and the effective sign/phase \cite{lecturenotes-2019} is defined as $s_k({\mathbf X})\equiv D_k({\mathbf X})/|D_k({\mathbf X})|$. The Metropolis procedure samples $|D_k({\mathbf X})|$ and the expectation value is obtained from the Monte Carlo samples as
\begin{equation}
\langle \hat{A}\rangle_{k} \approx \frac{\sum_{\mathbf X} s_k({\mathbf X}) A_k({\mathbf X})}{\sum_{\mathbf X} s_k({\mathbf X})}=\frac{\langle s_k A_k\rangle_{|D_k|}}{\langle s_k\rangle_{|D_k|}}\,,
\label{eq:MCMC_A_k2}
\end{equation}
where on the right-hand side, the averages are with respect to the Monte Carlo samples, $\{{\mathbf X}\}$, drawn from $|D_k({\mathbf X})|$. Any MCMC algorithm can now be applied, starting from an initial auxiliary-field configuration as given by ph-AFQMC after back propagation. This can be, for example, using standard Metropolis by proposing new fields one slice ${\mathbf x}^i$ at a time from $p({\mathbf x}^i)$. Alternatively, we can introduce a force bias \cite{HaoShi_Metropolis_force}, i.e., use Eq.~(\ref{eq:HS-shifted}) in place of Eq.~(\ref{eq:HS}) in defining $D(X)$ above \cite{lecturenotes-2019}, which amounts to considering a shifted probability density in the sampling. Any other acceleration scheme can be applied. Similarly, any improvement to the standard Metropolis approach in AFQMC can be applied.

In Eq.~(\ref{eq:Metrop-release}) the exact values of $\{W_k^{m+n}\}$ are not known. However, very good estimates of their values are provided by the ph-AFQMC calculations from which the initial paths are generated. This amounts to replacing the importance function in Eq.~(\ref{eq:future-wt-step1}) with an approximate version from the phaseless calculation. The simplest is to use the ``bare'' weight  $\overline{W}^{n+m}_k$, which corresponds to using $I_{\rm ph}$ (as given by Eq.~(\ref{eq:importance_function_CP})) for $I$. An improved approximation is to use the weight from path restoration (PRes) \cite{AFQMC_PathRestoration}. With PRes, we can restore one or both parts of the $\cos$ projection and the overall phase, either partially or wholly \cite{Siyuan-PW}. The latter is a natural choice for the present purpose since the ensuing Metropolis calculation will need the entire phase along the path. We find that not restoring the $\cos$ projection, which reduces noise, does not seem to affect the final systematic accuracy of the constraint release results.
 
Putting all of the above together, we implement Eq.~(\ref{eq:Metrop-release}) in our Metropolis constraint release as follows:
\begin{equation}
\langle \hat{A} \rangle \approx \frac{\sum_k\overline{W}^{m+n}_k \langle s_k A_k\rangle_{|D_k|}}{\sum_k \overline{W}^{m+n}_k \langle s_k\rangle_{|D_k|}}
\,.
\label{eq:Metrop-release2}
\end{equation}
The global signal-to-noise ratio for the Metropolis release constraint calculation is monitored through 
\begin{equation}
S = \Big|\sum_k \overline{W}_k^{m+n} \langle s_k\rangle_{|D_k|}\Big|/\sum_k \Big|\overline{W}_k^{m+n}\langle s_k\rangle_{|D_k|}\Big|\,.
\label{eq:def-S}
\end{equation}
It is interesting to compare this to using Eq.~(\ref{eq:Metrop-release}) directly: 
\begin{equation}
\langle \hat{A} \rangle \approx \frac{\sum_k\overline{W}^{m+n}_k s_k^0\, \langle \hat{A}\rangle_{k}}{\sum_k \overline{W}^{m+n}_k s_k^0}
\,,
\label{eq:Metrop-release-v2}
\end{equation}
where $\langle \hat{A}\rangle_{k}$ is from Eq.~(\ref{eq:MCMC_A_k2}) and $s_k^0$ denotes the phase of the initial path produced by ph-AFQMC and back-propagation for $|\phi_k^n\rangle$, after the restoration of the phase along the path. The difference between the above two formalisms is two-fold. First, the approach in Eq.~(\ref{eq:Metrop-release2}) includes the relative phase fluctuations in the ensuing Metropolis sampling of each $D_k({\mathbf X})$ in approximating $W^{m+n}_k$. And secondly, it weighs the contribution of each $\langle \hat{A}\rangle_{k}$ according to the overall weight of $D_k({\mathbf X})$ for the {\it finite number\/} of samples. We find that the two approaches yield indistinguishable results for large sample numbers in the Metropolis, but Eq.~(\ref{eq:Metrop-release2}) leads to faster convergence with respect to the number of sweeps in the release calculation (for a fixed $\beta$ and fixed population $\{ |\phi_k^n\rangle \}$).

We comment on the use of a multi-determinant $\langle \Psi_T|$ in the Metropolis algorithm. When  $\langle \Psi_T|$ has the special form of multiple determinants from configuration interaction, for example, from a complete active space (CAS) calculation (CASCI or CASSCF) or SHCI, there are acceleration algorithms \cite{Hao_Some_recent_developments, SHCI_AFQMC, transition_metals_Shee} for fast computations in AFQMC. However, they are formulated to the mixed estimator for computing 
objects such as $\langle \Psi_T|\hat O |\phi\rangle /\langle \Psi_T|\phi\rangle$, where $|\Psi_T\rangle$ is the multi-determinant CI trial wave function, $|\phi\rangle$ is an arbitrary Slater determinant and $\hat O$ is a one-body/two-body operator or single-particle propagator  $B(\mathbf{x})$. In the Metropolis algorithm, $\hat O$ is typically inserted along the path, not adjacent to $\langle \Psi_T|$. This imposes additional technical hurdles since the CI determinants, after being propagated by a general $B(\mathbf{x})$, cease being orthonormal to each other.
In our calculations, for overlaps, we apply the single-particle propagators $B(\mathbf{x})$ to the right (instead of to the left) to convert the computation to the desired form. We also apply 
tricks used in finite-temperature AFQMC calculations \cite{Yuan-Yao_FT_AFQMC} to save blocks of the product of a number of $B(\mathbf{x})$'s. For observables, we apply the method of Vitali {\it et.~al.\/} \cite{Vitali_Calculating_2019} to move $\hat O$ to the left, which also helps with the use of more general forms of $|\Psi_T\rangle$, such as pseudo-BCS \cite{Zxiao_2020} or Hartree-Fock-Bogoliubov \cite{Hao_HFB}. We believe there is considerable room for additional development for Metropolis release constraint to accelerate the use of multi-determinant trial wave functions as well as other aspects of the algorithm.

We should also note that under the Metropolis formalism of AFQMC (or many other quantum Monte Carlo), an infinite variance problem tends to arise \cite{Hao_infinite_variance}. Its effect in molecular systems under complex auxiliary-field transformations is not well characterized, especially in the presence of a phase problem which typically contributes a more dominant factor to the variance. We leave these issues to a future study as the focus in this work is to introduce the new approach, which, as shown below, significantly reduces the instability in constraint release. 

\subsection{\label{sec:level32} Algorithm and implementation}

We sketch a high-level outline of the algorithm to interface the ph-AFQMC with Metropolis constraint release. The scheme is illustrated in  
Fig.~\ref{Fig.AFQMC_Metropolis}.
\begin{enumerate}
    \item Choose a projection time $\beta$ for the release to be performed by MCMC, using a bra $\langle \Psi_T|$ and ket which is an importance-sampled phaseless 
    ground state to be obtained in step 2 below.
    \item Perform regular ph-AFQMC with one back-propagation cycle of $\beta'$ ($\ge\beta$), using a trial wave function $|\Psi_T\rangle$, and any initial population of $N$ walkers as appropriate. Equilibrate the calculation for a sufficient $\beta_{{\rm I}\rightarrow {\rm ph}}$. Store all the info for back-propagation, including the parent walker $\{|\phi_k^n\rangle\}$, the entire auxiliary-field path $\{{\mathbf x}_k^{i}\}$, the corresponding shifts $\{{\overline {\mathbf x}}_k^{i}\}$, and the final weight of the path $\{ \overline{W}_k^{m+n} \}$. 
    \item Use the back-propagation path for each $|\phi_k^n\rangle$ to initialize an MCMC calculation as given by Eq.~(\ref{eq:Metrop-release-est}). The MCMC can start from the input field configuration, especially if force bias is used in the sampling \cite{HaoShi_Metropolis_force}. If not, the MCMC only needs the initial walker $|\phi_k^n\rangle$ information and can start from random initial fields. 
    \item Run the desired algorithm of MCMC for each walker $k$ independently as a regular path-integral AFQMC with no constraint. If $\beta'>\beta$, a partial release can be run while freezing the left-most portion of the path of length $(\beta'-\beta)$. (See Sec.~\ref{sec:observable}.)
    \item Collect converged MCMC results and average them, following Eq.~(\ref{eq:Metrop-release2}). 
\end{enumerate}

The case of $\beta'=\beta$ is normal, full release. The case of $\beta'>\beta$ allows for a portion of the path on the left, illustrated by the region of $l$ time slices, 
to be frozen at the ph-AFQMC path; this provides interesting possibilities to slow down the onset of the phase problem, which are discussed further in the next section.  

All our results in this paper were obtained with the hybrid version \cite{Hao_Some_recent_developments} of ph-AFQMC. In the back-propagation, we have chosen a simple version of the path restoration \cite{AFQMC_PathRestoration}, including the overall phase throughout the path (of length $\beta'$), but without undoing the $\cos$ projection. 

In our implementation of the Metropolis sampling, we tested the schemes discussed in the previous section. For example, we tested the full force bias approach \cite{HaoShi_Metropolis_force} using dynamically updated forces as in ph-AFQMC. We also tested a simpler version, which keeps a ``constant'' force bias as in the original ph-AFQMC sampled path. In other words, we use the same force bias $\{{\overline {\mathbf x}}_k^{i}\}$ throughout the MCMC sampling. Mathematically this is correct as Eq.~(\ref{eq:HS-shifted}) holds for any force bias. Somewhat surprisingly, this approach still maintained quite high acceptance rates. Given that it is computationally much less costly than the full dynamic force bias, we adopted this approach for most of the results presented below. 

\subsection{\label{sec:level33} Additional discussions}

In this section, we comment on several technical aspects, including the computation of the energy variance, which can help monitor convergence of the constraint release, and additional convergence acceleration in the computation of observables that do not commute with $\hat H$. We have focused on ground-state computations throughout this paper, but here we will comment on the generalization of our approach to finite-$T$ AFQMC.

\subsubsection{\label{sec:variance} Computation of $\langle \hat H^2\rangle$ and energy variance}

Unlike in regular ph-AFQMC, which is stable and can be carried out for an arbitrarily long imaginary time, constraint release is intrinsically unstable as $\beta$ increases. In ph-AFQMC, we can rely on redundancy to ensure convergence with respect to $\beta_{{\rm I}\rightarrow {\rm ph}}$. In constraint release, we need to stop at the smallest $\beta$ which exceeds $\beta_{{\rm ph}\rightarrow 0}$, or when the signal-to-noise ratio is lost from the phase problem, whichever occurs first. It is, therefore, very valuable to have better detection of convergence. The exact ground-state energy is not known, of course. The energy variance, however, vanishes in the limit of full constraint release. Thus the variance
$$\sigma^2= \langle \hat{H}^2\rangle - \langle \hat{H}\rangle^2\,$$
as a function of release time $\beta$ can provide valuable information to help gauge convergence.

The measurement of $\langle \hat H^2\rangle$ is computationally costly since, in the most general case, it can formally scale as $N^8$  with basis set or system size. Instead of direct computation, we use
$$\hat{H}^2=\hat{H}\, \lim_{\lambda \rightarrow 0} \frac{e^{\lambda \hat{H}}-e^{-\lambda \hat{H}}}{2\lambda}.$$
We can thus compute 
\begin{equation}
\sigma^2=\lim_{\lambda \rightarrow 0} \biggl[ \Big\langle H \frac{e^{\lambda H}-e^{-\lambda H}}{2\lambda} \Big\rangle - \langle H \rangle \Big\langle \frac{e^{\lambda H}-e^{-\lambda H}}{2\lambda} \Big\rangle \biggr]\,.
\label{eq:MRC-var}
\end{equation}
We decompose the $e^{\lambda \hat{H}}$ term by the transformation in Eq.~(\ref{eq:HS}) and sample it by Monte Carlo. Note that for a small finite $\lambda$, the Trotter error is canceled to the third order in the expression above. 

In the Metropolis framework, the above scheme is conveniently embedded if we choose $\lambda=\tau$. For example, 
\begin{equation}
\begin{aligned}
    \langle H e^{-\tau H} \rangle& = \biggl\langle \frac{\langle \Psi_T|He^{-\tau H}|\phi^{n+m}\rangle}{\langle \Psi_T|\phi^{n+m}\rangle}\biggr\rangle_{\mathbf X}\\
    &=\Biggl\langle \biggl\langle 
    \frac{\langle \Psi_T|\phi^{n+m+1}\rangle}{\langle \Psi_T|\phi^{n+m}\rangle} \frac{\langle \Psi_T|H|\phi^{n+m+1}\rangle}{\langle \Psi_T|\phi^{n+m+1}\rangle}\biggr\rangle_{\mathbf x'}\Biggr\rangle_{\mathbf X}\,,
\end{aligned}
\end{equation}
where the walker $|\phi^{n+m+1}\rangle$ is obtained by propagating $|\phi^{n+m}\rangle$ for another time step via the use of an extra auxiliary field ${\mathbf x'}$ in the spirit of the bridge link idea \cite{Hao_infinite_variance}. Similarly, $\langle H e^{\tau H} \rangle$ can be computed. Applying antithetic variate, we pair the sampling of ${\mathbf x'}$ with $-{\mathbf x'}$ to reduce fluctuations in the above computation of the variance.

\subsubsection{\label{sec:observable} Observables and frozen paths}

To evaluate an observable $\hat{A}$ which does not commute with $\hat{H}$, we insert $\hat{A}$ in the middle of Metropolis chain in Eq.~(\ref{eq:Metrop-release-est}): 
\begin{equation}
\langle \hat{A} \rangle_k \approx \frac{\langle \Psi_T|\,e^{-(m-{\overline n})\tau \hat{H}}\,\hat{A}\,
e^{-{\overline n}\tau \hat{H}}\,
|\phi^{n}_k \rangle}{\langle \Psi_T|\,e^{-m\tau \hat{H}}\,|\phi^{n}_k\rangle}
\,,
\label{eq:Metrop-release-est-middle}
\end{equation}
where the position of ${\overline n}$ can be adjusted.
Since ph-AFQMC yields a wave function  $|\Psi^n\rangle$ which is close to the true ground state $|\Psi_0\rangle$, the number ${\overline n}$ can be small, so the optimal location for measuring $\langle A\rangle$ is near the right end of the path. In practice, we measure for a number of different locations and $m$ to help gauge convergence, as illustrated in the next section. For the special case of the energy (or an observable which commutes with $\hat H$, as mentioned in Sec.~\ref{sec:level31}, any choice of ${\overline n}$ is correct, including ${\overline n}=0$ and ${\overline n}=m$ (the mixed estimator).

For pure estimators, we can freeze a portion of paths in the Metropolis sampling to further improve efficiency. In Fig.~\ref{Fig.AFQMC_Metropolis}, this is illustrated by the ``Frozen" region, where the auxiliary fields are fixed during the MCMC at the input values from ph-AFQMC. Effectively this can be thought of as replacing $\langle \Psi_T|$ in Eq.~(\ref{eq:Metrop-release-est-middle}) above by $\langle \Psi_T|e^{-l\tau {\hat H'_{\rm BP}}}$, where we have used ${\hat H'_{\rm BP}}$ to denote the effective action from the Hamiltonian, but under two modifications: the constraint in the forward direction (indicated by the prime on $\hat H$) and under back propagation. As we show in Sec.~\ref{sec:level31}, this scheme can reduce the length of the imaginary time needed in constraint release to reach a desired level of convergence. This can either yield greater statistical accuracy (larger average phase for a given release time) or improve convergence by allowing a longer release time.

\subsubsection{\label{sec:level333} Constraint release in finite-$T$ AFQMC}

We have presented our formalism and discussions all in the context of ground-state AFQMC. Finite-temperature constrained path AFQMC
\cite{zhang1999finite,Yuan-Yao_FT_AFQMC} or the phaseless counterpart \cite{lecturenotes-2019} is also done with a branching random walk, except that the random walk has a finite length $\beta$. No $|\Psi_I\rangle$ or $|\Psi_T\rangle$ is involved. There is a trial density matrix which imposes the constraint. The finite-$T$ calculation, as in the ground-state algorithm, has a population of $N_{\mathbf x}$ random walkers. The ones which survive the branching processes through the final step (defined by inverse temperature $\beta$) yield a set of $N_{\mathbf x}$ paths, which can have different weights depending on the particular population control algorithm employed \cite{Yuan-Yao_FT_AFQMC}.
For each walker, at the completion of the random walk, measurements are performed at different imaginary-time locations, and the results are averaged, effectively ``rotating'' the path to help restore time translational invariance. This is a further simplification from the ground-state method, where the location of the measurement matters, as discussed in Sec.~\ref{sec:observable}. 
In the finite-$T$ method, the final results is a weighted average of the $N_{\mathbf x}$ walkers (or paths) with respect to their final weights. Repeats are carried out as needed, of the entire $m$-step random walk process with the population of $N_{\mathbf x}$ walkers.

The Metropolis release algorithm we have described can be generalized straightforwardly to finite-$T$ AFQMC. After the fixed-length random walk has been completed, each of the $N_{\mathbf x}$ walkers defines a full path. This has the same role as one full path from the back-propagation in ground-state ph-AFQMC. We can thus simply perform unconstrained MCMC sampling with the constrained walker for full or partial path serving as the initial configuration. The weights of the walkers serve the same role as the ``future weights'' $\{\overline{W}_k^{m+n}\}$ in the ground-state method. Each fully converged MCMC sampling for a full path would lead to a full finite-$T$ calculation, but the results as a function of the Monte Carlo time and the length of the MCMC sampled path will give the results of constraint release. That is, the convergence or equilibration of the MCMC release constraint connects the constrained result to the unconstrained result continuously.

\section{\label{sec:level4}Results}

In this section, we apply the algorithm outlined in the previous section to a number of atoms and molecules. All calculations are for the ground state and in a small but realistic basis set where accurate reference data exists. We chose a range of systems that present challenging and interesting cases. These include certain main group molecules for which a single determinant $|\Psi_T\rangle$ produces poor results in ph-AFQMC. They also include more correlated systems: bond breaking in N$_2$ and transition metal containing atoms and molecules, where we test multiple-determinant $|\Psi_T\rangle$'s as is typically done, but also examine worst-case scenarios with single-determinant $|\Psi_T\rangle$. In all cases, the Metropolis release constraint removes the residual error from the phaseless constraint, and achieves chemical accuracy. 

With these examples, we aim to both test the algorithm and study its behavior. We investigate the stability of the constraint release and consider the convergence criteria, including the use of the energy variances as described in Sec.~\ref{sec:variance}. Our ph-AFQMC calculations follow standard practices \cite{WIREs_Mario,Hao_Some_recent_developments}. Some of the main-group calculations are all-electron, but the high-accuracy extrapolated ab-initio thermochemistry (HEAT) set \cite{HEAT_set} used frozen-core so as to compare directly with existing data \cite{sukurma2023_benchmark}. The transition metal systems remove the core electrons and use effective core potentials (ECP) \cite{pseudopotentials_1,pseudopotentials_2,pseudopotentials_3} following an earlier benchmark study \cite{simons_material_2020}, which provides results for direct comparison. The HF and complete active space self-consistent field (CASSCF) calculations were performed with PySCF \cite{pyscf}, which were imported into AFQMC (with truncation in the case of CASSCF, to reduce $|\Psi_T\rangle$ to a modest multi-determinant wave function, some with ${\mathcal O}(1000)$ determinants but most with much less). Conservative cutoffs for the Cholesky decomposition were chosen. We used a $\tau = 0.01/\textup{Ha}$, and verified that any residual error in the Metropolis is much less than the statistical error. Below we first describe our test on main-group systems, including ``worse" cases in the HEAT set and a study of frozen paths in the computation of observables. We then present results on more correlated systems, first bond-breaking in N$_2$ followed by transition metal systems.

\subsection{\label{sec:level41}Main group molecules}

\subsubsection{\label{sec:level411}Illustration with the HEAT set }

\begin{figure}[htbp] 
\includegraphics[scale = 0.2]{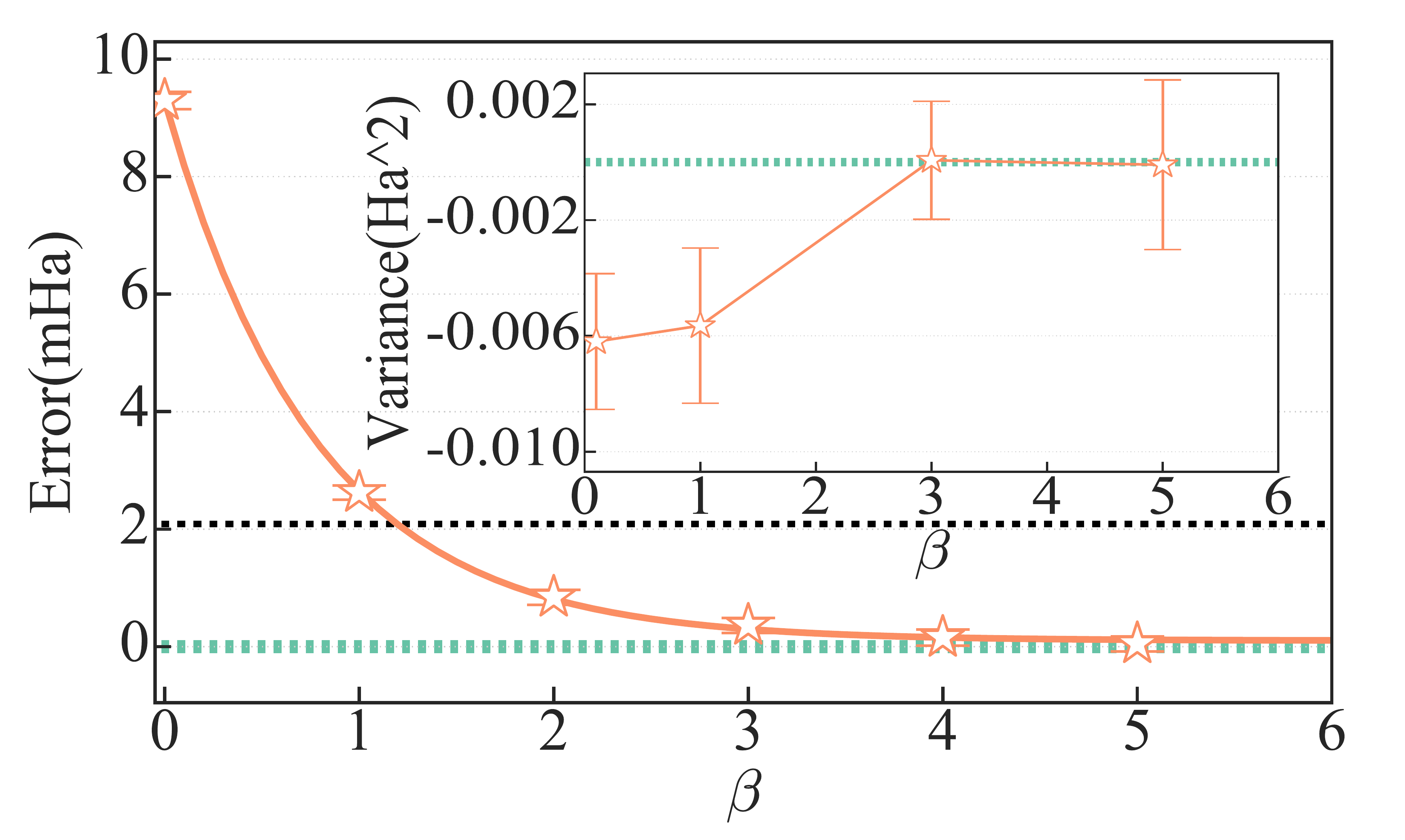}
\caption{
Metropolis release constraint calculation in O$_2$. The error in the computed total energy with respect to the exact result (from CCSDTQP in Ref.~ \cite{sukurma2023_benchmark})
is shown as a function of release imaginary-time $\beta$. A single Slater determinant from UHF is used as $|\Psi_T\rangle$. The result at $\beta=0$ is ph-AFQMC. CCSD(T) result is shown for reference (black dotted line). Statistical error bars are shown but are smaller than symbol size; the solid line is to guide the eye. The inset shows the computed energy variance, which converges to zero with $\beta$.
} 
\label{Fig.Metro_AFQMC_O2} 
\end{figure}

We will first illustrate our constraint release approach using a set of small molecules in the HEAT set. A recent paper \cite{sukurma2023_benchmark} reported a new Fortran implementation of the ph-AFQMC method and interface with VASP \cite{VASP}. Benchmark results for  26 molecules in the HEAT set were presented using single determinant trial wave functions from RHF or UHF. A number of molecules were highlighted for which this approach did not reach chemical accuracy for the total ground-state energy. This can, of course, be remedied straightforwardly by the use of compact multi-determinant trial wave functions \cite{AFQMC_bondBreaking,transition_metals_Shee}, but here we will restrict to the use of single determinant $|\Psi_T\rangle$, and use these molecules as test cases to show the behavior of the constraint release algorithm. 

To allow direct comparisons with Ref.~\cite{sukurma2023_benchmark}, we adopt the same frozen-core approximation, basis set (cc-pVDZ), geometries \cite{HEAT_geometry}, and (conservative) threshold of $10^{-8}$ for truncating the Cholesky vectors from the molecular electron repulsion integrals (ERIs). As shown below, the computed ph-AFQMC total energies are in reasonable agreement, which gives us a comparable starting point for the Metropolis release constraint. For the ph-AFQMC calculations, we also use a small time step of 0.002\,Ha$^{-1}$. We note that the constraint release or free projection has quadratic Trotter error in contrast with ph-AFQMC, where it is typically linear \cite{hybrid_phaseless,WIREs_Mario}. Thus larger  $\tau$ values can be adopted in the Metropolis calculations, as mentioned earlier. We have verified that the residual Trotter error is smaller than our statistical error bar in all our Metropolis release constraint calculations. 

Figure~\ref{Fig.Metro_AFQMC_O2} shows an example of the  Metropolis release constraint calculation in O$_2$, which was seen to have the largest phaseless constraint error (with a single Slater determinant  $|\Psi_T\rangle$) among the HEAT set: $\sim 9$\,mHa \cite{sukurma2023_benchmark}. This is consistent with the result from the original implementation of ph-AFQMC for Gaussian basis sets \cite{AFQMC_mol-Gau-oririnal-Alsaidi2006}. The Metropolis release constraint reduces this error to less than $1$\,mHa with a projection time of $\beta=2$. The computation can be comfortably continued to large $\beta$, in this case, to reach convergence with a resolution well beyond $0.1$\,mHa. We can also compute the energy variance following the discussion in Sec.~\ref{sec:variance}. The result is shown in the inset. Note that the estimator in Eq.~(\ref{eq:MCR-me}) is not variational since the bra and ket in Eq.~(\ref{eq:MCR-me}), $\langle \Psi_T|$ and $|\Psi^n\rangle$, are different. Even in the case $|\Psi_T\rangle=|\Psi_I\rangle$, ph-AFQMC or constrained path AFQMC is not variational \cite{issue-CPMC-1999,lecturenotes-2019}. The the energy does not necessarily approach the exact result from above, as we will see below. Furthermore, the variance does not have to be positive. However, the variance does vanish when the exact wave function is reached. As we see in the figure, the magnitude of the variance approaches zero. This gives a second indicator to help gauge convergence, especially when the energy convergence is less clear-cut than in the present case. 

\begin{figure}[htbp] 
\includegraphics[scale = 0.2]{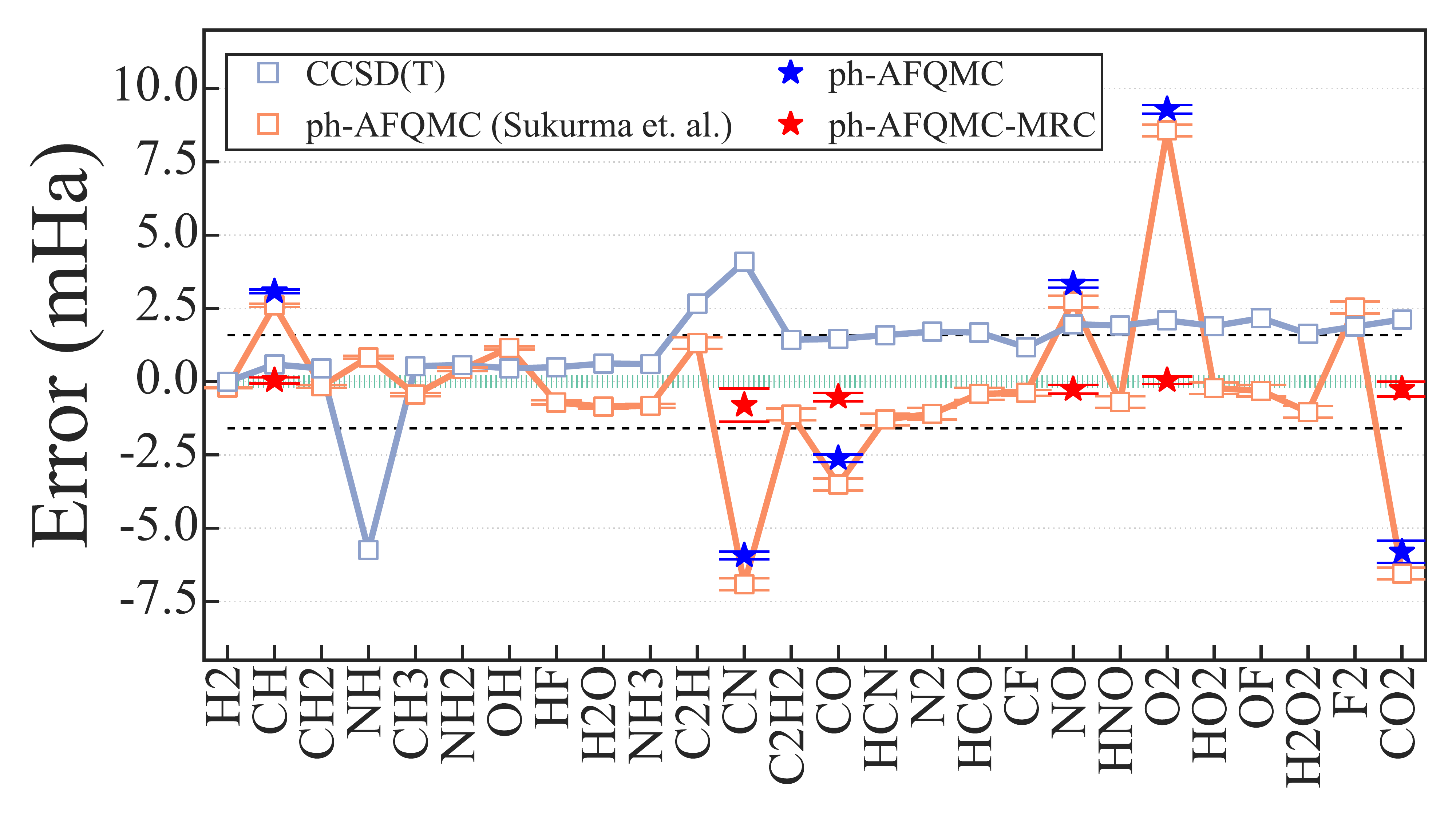}
\caption{Metropolis release constraint in the HEAT set. All calculations use single determinant $|\Psi_T\rangle$. The error in the computed total ground-state energy is shown with respect to the CCSDTQP result, which is exactly well within our statistical error bar. Blue stars are ph-AFQMC results, while red stars show the corresponding Metropolis release constraint result. All other data are from Sukurma {\it et.~al.\/}~\cite{sukurma2023_benchmark}, which are reproduced for all 26 molecules for reference. Metropolis release constraint systematically restores chemical accuracy.
} 
\label{Fig.HEAT} 
\end{figure}

We carried out  Metropolis release constraint calculations for all the molecules identified as outliers in Ref.~\cite{sukurma2023_benchmark}: CH, CN, CO, NO, O$_2$,  and CO$_2$. All calculations used single determinant $|\Psi_T\rangle$ from Hartree-Fock. The results for CH use RHF  while others use UHF for $|\Psi_T\rangle$. To be consistent with Ref.~\cite{sukurma2023_benchmark}, we use the cc-pVDZ basis set and apply a frozen-core approximation ($1s$, except for H). In Fig~\ref{Fig.HEAT}, we plot the result together with the ph-AFQMC and reference data reproduced from  Ref.~\cite{sukurma2023_benchmark}. In all cases, the Metropolis release constraint calculations converged to the reference result (CCSDTQP) to well within chemical accuracy. The computation time is modest, comparable to that of the corresponding ph-AFQMC. For example, O$_2$ takes about 5$\times$ their ph-AFQMC time at $\beta=4$, and CN takes about 16$\times$ their ph-AFQMC time at the $\beta=15$, which is the longest $\beta$ to converge in these Heat tests.

\subsubsection{\label{sec:level412}Computation of observables and frozen paths}

\begin{figure}[tbp]
\includegraphics[scale = 0.4]{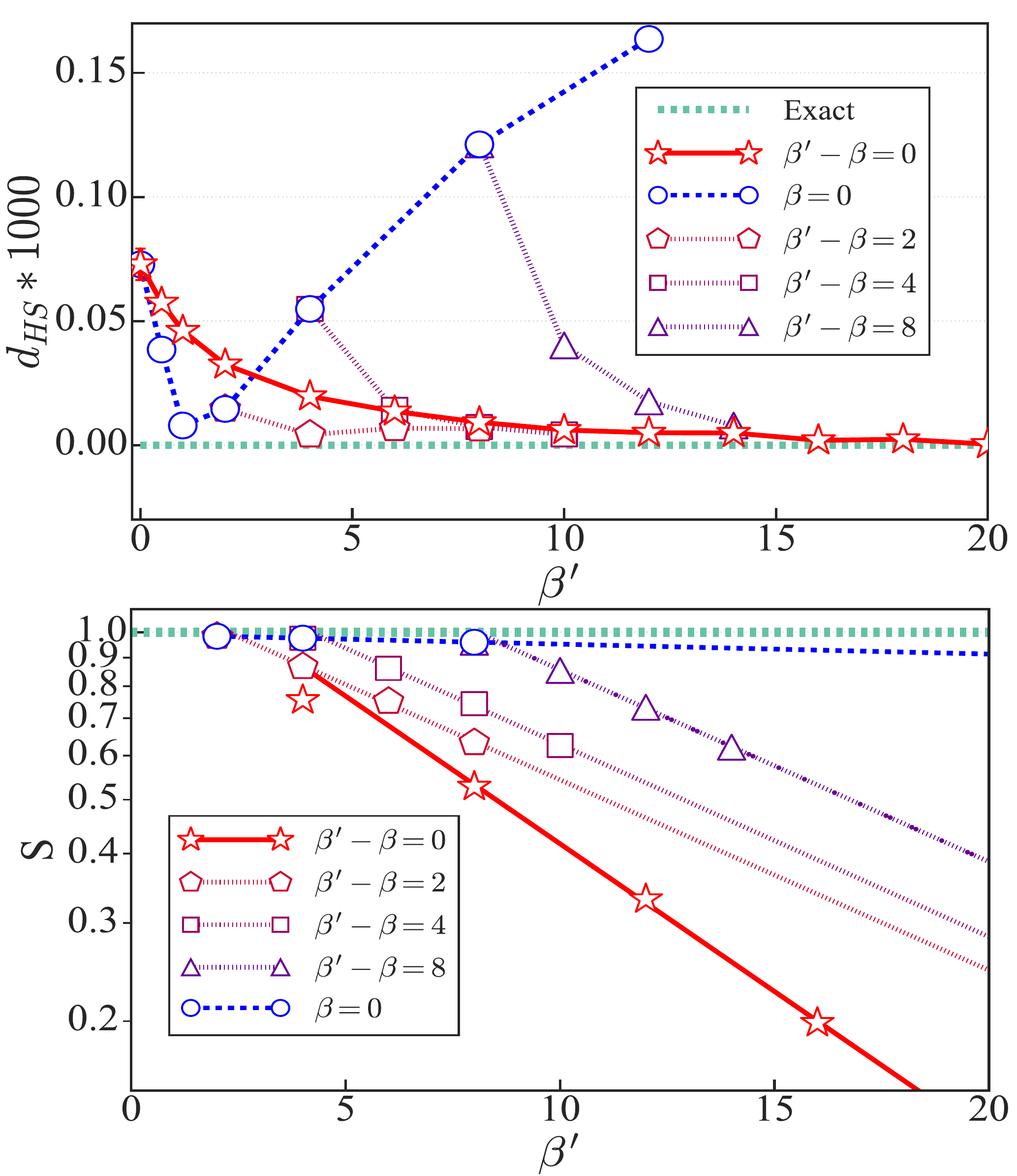}  
\caption{Computation of the one-body density matrix (1RDM), and effect of frozen paths in the Metropolis release constraint in CH+ at bond-length $R=1.146$\,\AA. Calculations here are all-electron, using the cc-pVDZ basis. A single Slater determinant from RHF is used as $|\Psi_T\rangle$. The top panel shows the normalized Hilbert-Schmidt distance $d_{HS}$ (Eq.~(\ref{equ:Hilbert-Schmidt})) of the computed 1RDM from the exact result, as a function of total path length $\beta'$ and freezing portions of the path of varying lengths $\beta'-\beta$ next to $\langle\Psi_T|$, as illustrated in Fig.~\ref{Fig.AFQMC_Metropolis}. The bottom panel shows the evolution of the phase problem, as measured by the signal-to-noise ratio $S$ (Eq.~(\ref{eq:def-S})). 
} 
\label{Fig.CH-BP} 
\end{figure}

In this section, we illustrate the computation of general observables which do not commute with the Hamiltonian. As discussed in Sec.~\ref{sec:observable}, this can be done by 
making measurements away from the two ends of the paths. In the test below, we measure at the middle of the Metropolis path, i.e., with $\bar{n}=m/2$ (see Fig.~\ref{Fig.AFQMC_Metropolis}).
We use, as an example, the CH$^+$ molecule at bondlength $R=1.146$\,\AA, and compute the one-body density matrix (1RDM) $O_{i,j}= \langle \hat{c}^\dagger_i\hat{c}_j\rangle$. For this system (cc-pVDZ basis), the exact 1RDM can be obtained by full Configuration Interaction (FCI), and we quantify the error using the normalized Hilbert-Schmidt distance from the FCI results:
\begin{equation}
d_{HS}= \sqrt{\textup{Tr}[(O-O_{\textup{FCI}})^2]}/{\mathcal N}\,,
\label{equ:Hilbert-Schmidt}
\end{equation}
where ${\mathcal N}$ denotes the number of matrix elements in $O$.

The test results are presented in Fig.~\ref{Fig.CH-BP} . Our calculation used a single Slater determinant $|\Psi_T\rangle$ from RHF. As shown in the top panel, the usual mixed results (i.e. $\beta=0$ and $\beta'-\beta=0$) in a noticeable bias, which is systematically removed by the Metropolis release constraint as $\beta$ is increased. The signal-to-noise ratio $S$ remains $\sim 0.2$ at $\beta= 16$\,Ha$^{-1}$. The ph-AFQMC ground-state energy has a constraint error of $-3.2\pm0.3$\,mHa, reduced to $-0.3\pm0.5$\,mHa in the Metropolis release constraint at $\beta=6$.

We also investigate the effect of freezing a portion of the paths in the Metropolis release constraint. As illustrated in Fig.~\ref{Fig.AFQMC_Metropolis}, we can keep a portion of the path at the left side (i.e. $l\tau=\beta'-\beta$) fixed to reduce the bias. We introduce the improvement of the frozen path by proceeding Metropolis release constraint with a different frozen path $\beta'-\beta$. One extreme case is $\beta=0$ where, as $\beta'$ increases, $d_{HS}$ decreases at the beginning and diverges eventually with negligible $S$. In this case, $d_{HS}$ suggests the quality of left walker $\langle \Psi_T|e^{-l\tau {\hat H'_{\rm BP}}}$ (See Sec.~\ref{sec:observable} ). As propagation continues and with no clear criteria to stop at minimal, the left walker is finally discarded by accumulated bias in phaseless approximation. One optimized way is proceeding with Metropolis release constraint with an appropriate frozen path (i.e. $\beta'-\beta=2,4,8$ ). As $\beta$ increases, their trajectories connect two lines (i.e.  $\beta=0$ and $\beta'-\beta=0$). The Metropolis release constraint with a better left walker (i.e. less $d_{HS}$) reaches a certain accuracy with less $\beta$ while maintaining a higher signal-to-noise ratio. For example, the optimized Metropolis release constraint run for this test is with frozen path $\beta'-\beta=2$, which converges at $\beta=2$ with $S \sim 0.85$.

\subsection{\label{sec:level42}Test in more strongly correlated systems}

\subsubsection{\label{sec:level421}Bond breaking in N$_2$}

\begin{figure}[htbp]
\includegraphics[scale = 0.4]{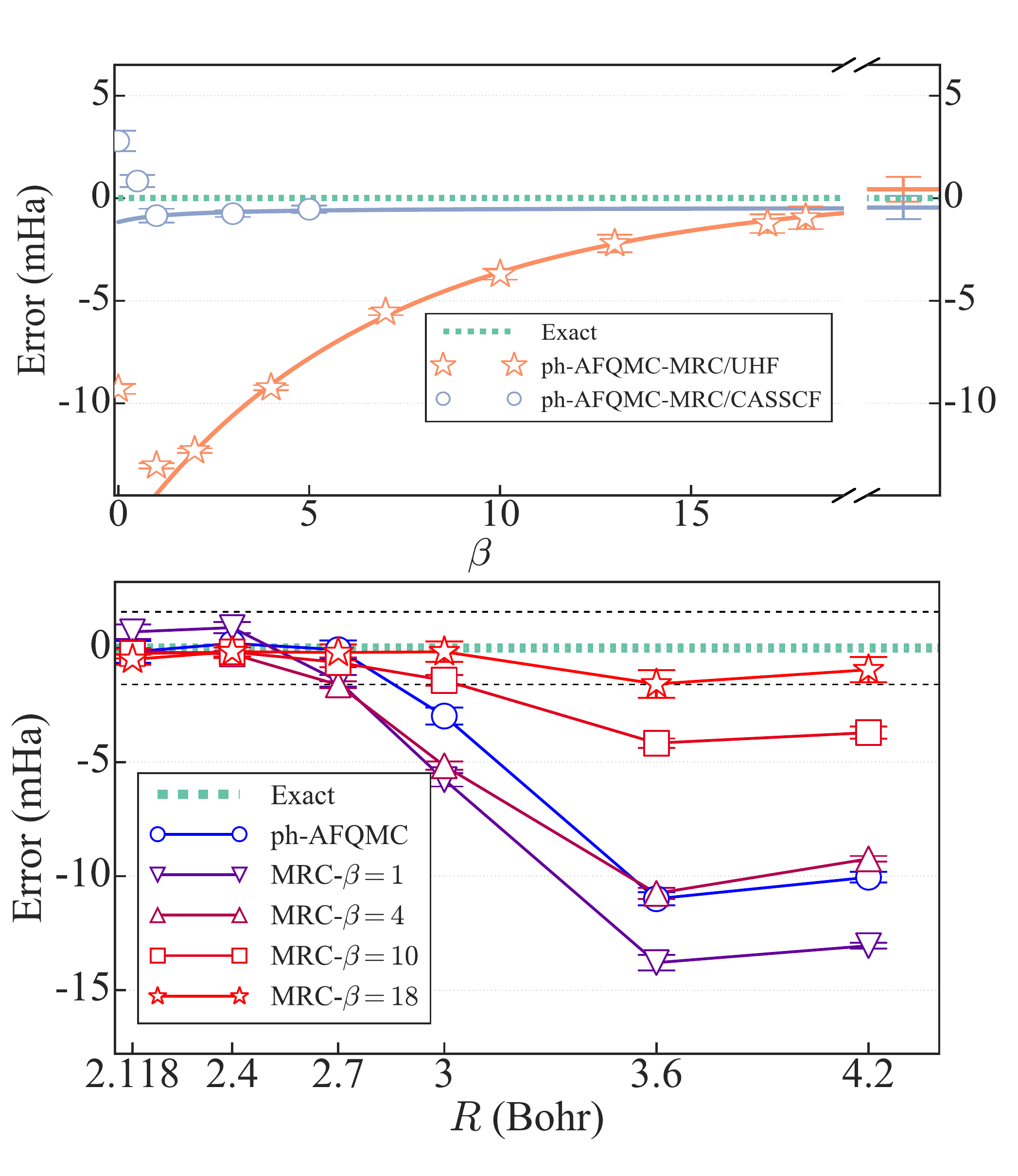} 
\caption{
Bond-breaking in N$_2$. All-electron calculations are performed, using the cc-pvDZ basis. The reference exact result is from Ref.~\cite{N2_bondbreaking_DMRG}. The top panel is for 
$R=4.2\,$Bohr. The computed energy is plotted versus constraint release time $\beta$, for two different choices of the trial wave function $|\Psi_T\rangle$: UHF and a truncated 
CASSCF(6e,12o). The solid lines show a fit of a exponential function, to guide eye; the asymptotic ($\beta\rightarrow \infty$) values from the fit are shown at the far right. The bottom panel shows the error for entire range of $R$, with different curves showing the progression of the constraint release. For these calculations, single-determinant UHF (RHF) $|\Psi_T\rangle$'s are used. 
} 
\label{Fig.N2}
\end{figure}

N$_2$ bond breaking is often used as a testbed for correlated methods, with the presence of a triple bond and the abundance of reference data. In this subsection, we test the Metropolis release constraint approach in this system. A systematic study of N$_2$ bond breaking with ph-AFQMC was performed during the early development of the method \cite{AFQMC_bondBreaking}. Results using compact multi-determinant wave functions as $|\Psi_T\rangle$, truncated from CASSCF, are very accurate, with maximum error of several mHa at the largest bondlength of $4.2$\,Bohr, while results from UHF $|\Psi_T\rangle$ have errors up to $\sim 9$\,mHa. We will therefore focus on the use of single-determinant $|\Psi_T\rangle$'s in the present study, although we will show one example of a truncated CASSCF $|\Psi_T\rangle$ for comparison. 

Results are presented in Fig.~\ref{Fig.N2}. The top panel focuses on the case of bondlength $R=4.2$\,Bohr. Two sets of Metropolis release constraint calculations are shown, one with UHF  
and the other with truncated CASSCF as $|\Psi_T\rangle$. The CASSCF calculation was performed with $6$ active electrons and $12$ CAS orbitals, with a weight cutoff of $0.01$, which is implemented by removing determinants with the smallest weights where the sum of the total removed probability is less than $0.01$. These choices are consistent with Ref.~\cite{AFQMC_bondBreaking}. In the Metropolis release constraint calculations using this multi-determinant $|\Psi_T\rangle$, we adopt the fast-update algorithm \cite{Hao_Some_recent_developments} as discussed in Sec.~\ref{sec:level31}. Both sets of calculations are seen to converge to the exact answer, with the truncated CASSCF $|\Psi_T\rangle$ requiring much shorter $\beta$ to reach chemical accuracy. As mentioned in Sec.~\ref{sec:level411}, the computed ground-state energy is not variational. In the figure, we show an exponential fit of $E(\beta)$ versus $\beta$ for each calculation, where  $E(\beta)$ denotes the computed ground-state energy (shown as the difference from the reference energy) as a function of projection time. A single exponential is an ansatz based on the assumption that, when sufficiently close to convergence to the exact energy, the imaginary-time projection is dominated by one higher energy state (hence one gap), and $E(\beta)=E_0+ae^{-\beta\Delta}$. We see that the fit quality is very good for sufficiently large $\beta$, although non-monotonic convergence is seen at smaller $\beta$, which is related to the non-variational nature mentioned above.

In the bottom panel, we investigate the behavior of the  Metropolis release constraint at different strengths of the electron correlation by computing the entire potential energy curve (PEC).
Using compact multi-determinant $|\Psi_T\rangle$'s (for which the $R=4.2$\,Bohr case represents one of the worst situations) leads to chemical accuracy across the PEC for a projection time $\beta<1$. Instead, we use single reference $|\Psi_T\rangle$. The error of ph-AFQMC (with UHF/RHF $|\Psi_T\rangle$ ) is larger than needed for chemical accuracy for bond-length $\geq 3$ Bohr. The progression of the Metropolis release constraint results is shown for four values of projection time $\beta$. The non-monotonic convergence mentioned above is seen at smaller $\beta$, with the energy first getting worse at $\beta=1$ and $4$ for intermediate bondlengths. Beyond $\beta\sim 4$, the error is monotonically reduced, and, by $\beta \sim 20$, the PEC from Metropolis release constraint converges to the exact results within chemical accuracy.

\subsubsection{\label{sec:level422}Transition metal atoms and molecules}

We next consider a selected number of transition metal atoms and molecules, which are among the more challenging strongly correlated systems and which provide a complementary set of tests to the above. In particular, we apply Metropolis release constraint to Cu+, testing both RHF and CASSCF trial wave functions, and then to CrO and VO, using only single Slater determinant $|\Psi_T\rangle$'s. For the molecules, the RHF/UHF trial wave functions lead to significantly larger errors from the phaseless constraint than a compact truncated CASSCF $|\Psi_T\rangle$ as adopted in Ref.~\cite{simons_material_2020}; however, we take the worst case scenario here for the purpose of testing. Finally, we apply the method to FeO (using a multi-determinant $|\Psi_T\rangle$) and examine the calculation of the energy variance in a strongly correlated system.

For all systems, we use the same ECP's \cite{pseudopotentials_1,pseudopotentials_2,pseudopotentials_3} and the VDZ basis set as in Ref.~\cite{simons_material_2020} to facilitate direct comparisons. We performed CASSCF with $10$ active electrons and $19$ CAS orbitals for Cu+, and $12$ active electrons and $9$ CAS orbitals for FeO. To obtain a truncated multi-determinant $|\Psi_T\rangle$, a weight cutoff of $0.01$ was applied in Cu+ and $0.001$ in FeO CASSCF. For VO and FeO, we use RHF $|\Psi_I\rangle$ for spin filtration \cite{wirawan-F2-spin-filtration}. Other calculations were performed with the same $|\Psi_I\rangle$ and $|\Psi_T\rangle$. 

\begin{figure}[htbp]
\includegraphics[scale = 0.205]{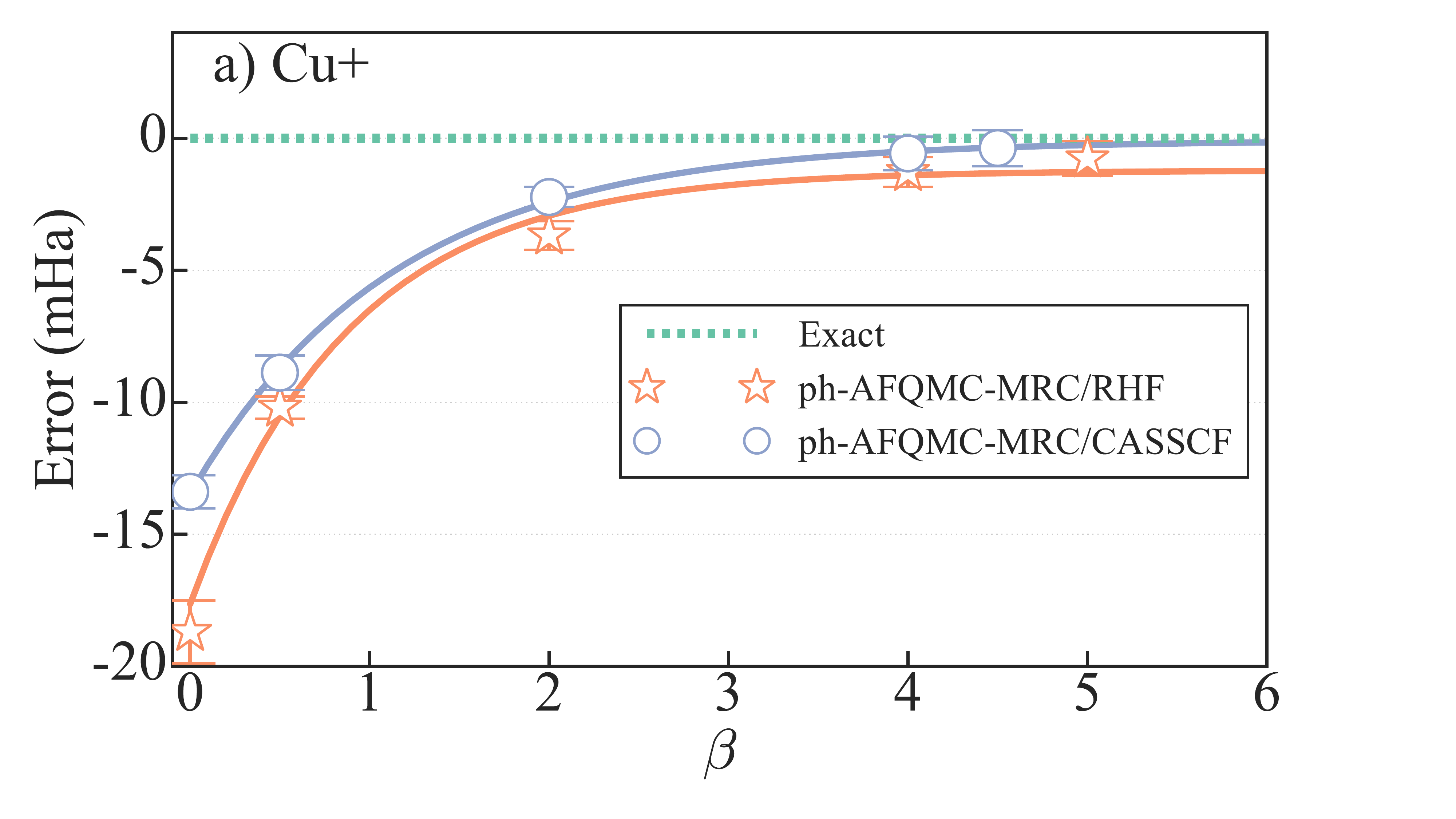}
\includegraphics[scale = 0.2]{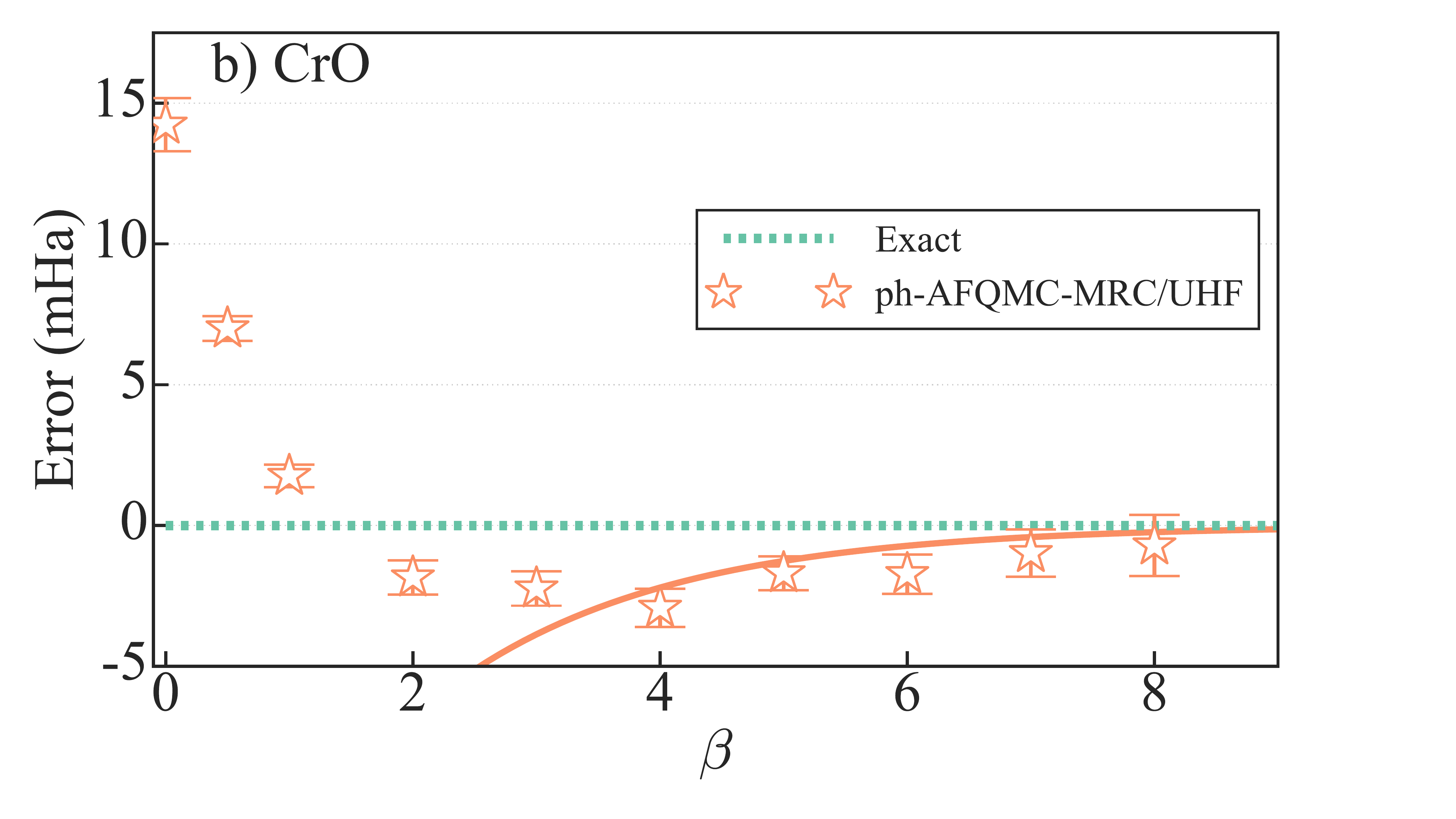}
\includegraphics[scale = 0.2]{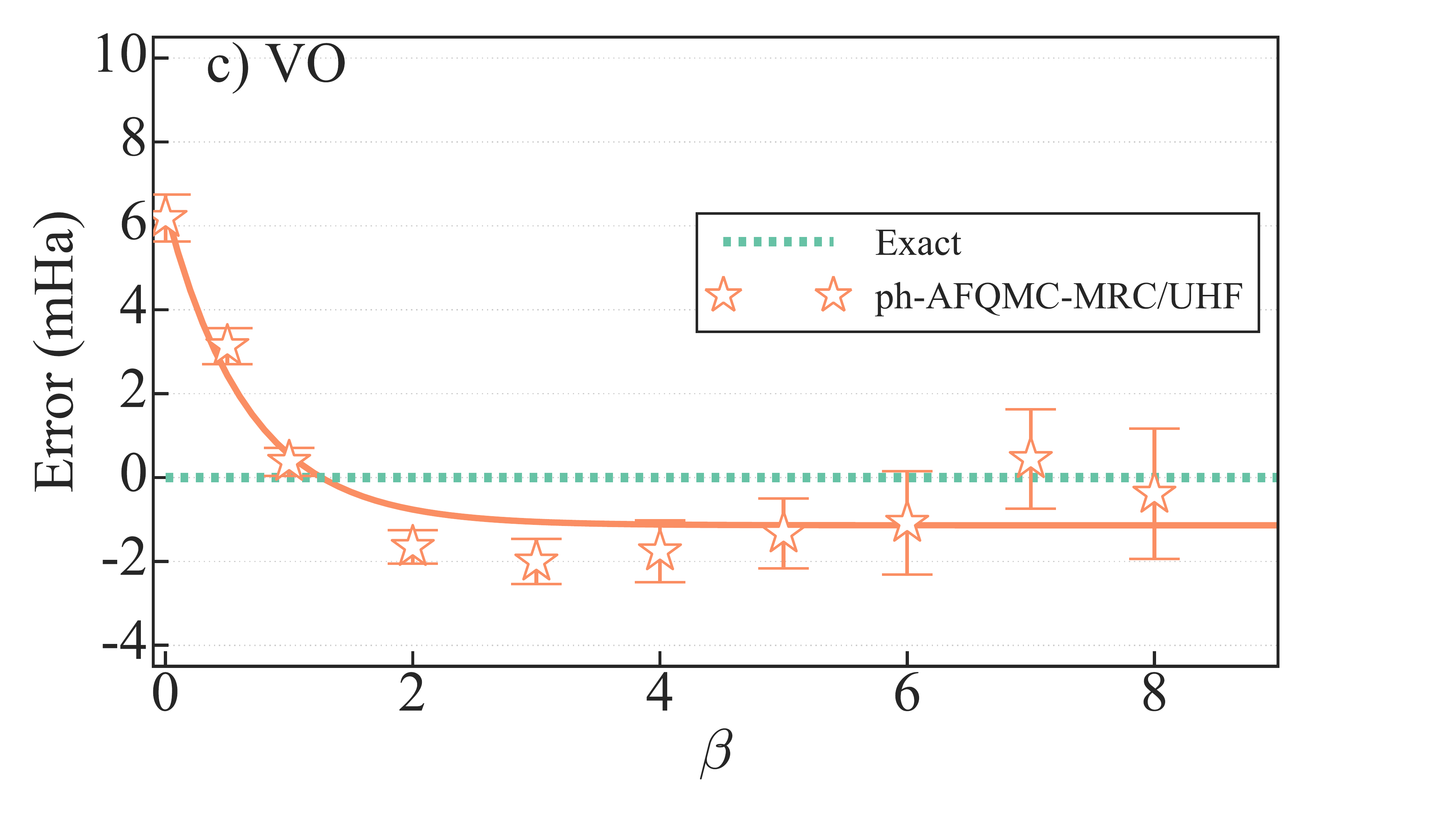}
\caption{Metropolis release constraint in Cu$^+$, CrO, and VO, three examples with unusually large ph-AFQMC errors from previous benchmark studies \cite{simons_material_2020,James-IP}. Results are shown for both RHF and CASSCF($10\textup{e}$, $19\textup{o}$) $|\Psi_T\rangle$ in Cu$^+$, and UHF $|\Psi_T\rangle$'s in CrO and VO. The solid lines show single exponential fits for larger $\beta$ as guide to the eye.  
} 
\label{Fig.Transition_metal} 
\end{figure}

\begin{figure}[htbp]
\includegraphics[scale = 0.2]{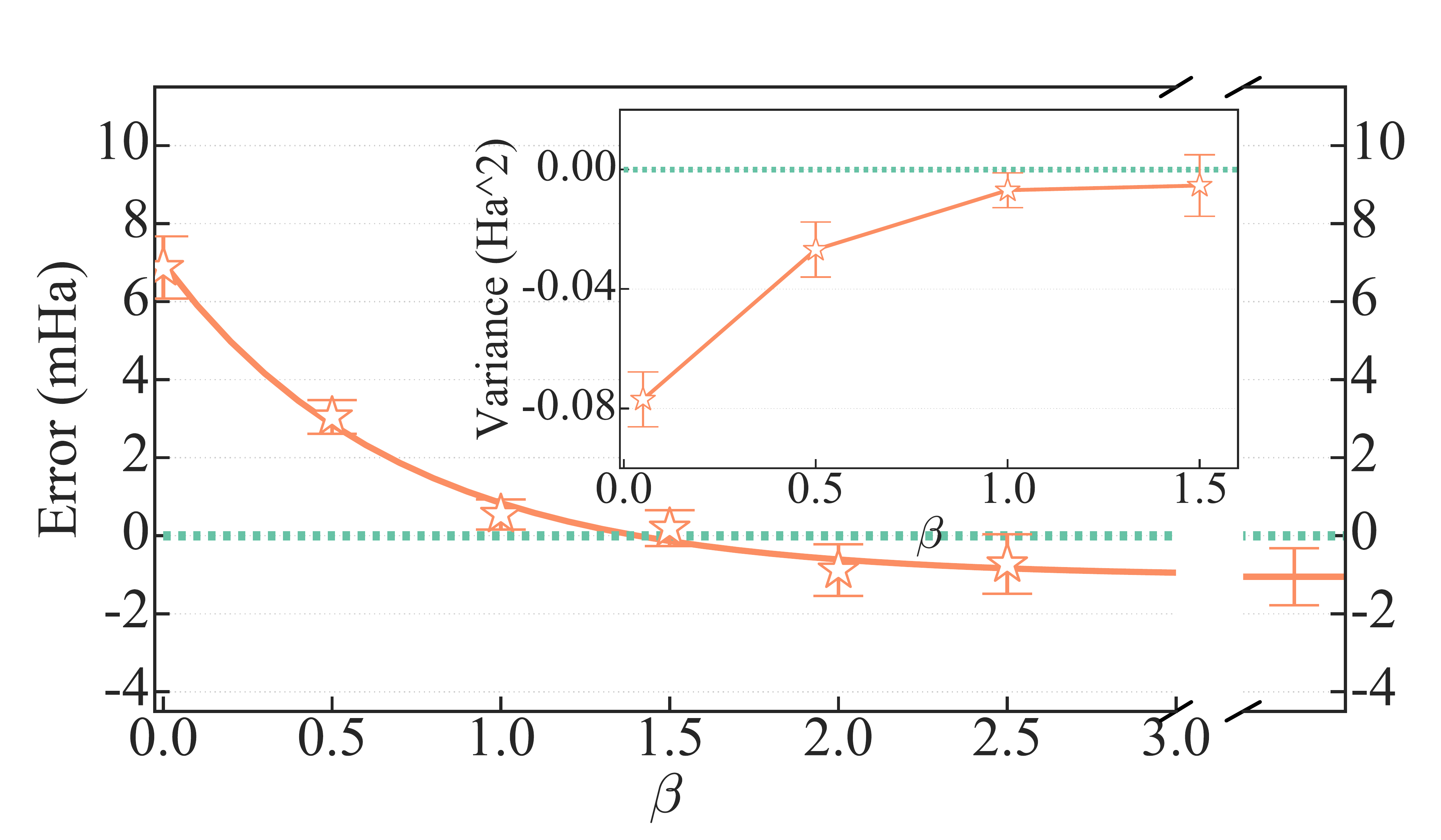}  
\caption{
Metropolis release constraint in FeO. The computed total energy is plotted as a function of release imaginary-time $\beta$, with respect to the near-exact reference result in Ref.~\cite{simons_material_2020}. Results are shown for a CASSCF($12\textup{e}$, $9\textup{o}$) $|\Psi_T\rangle$ and RHF $|\Psi_I\rangle$. With increasing projection time in the Metropolis release constraint, the initial ph-AFQMC bias of $\sim6.5$\,mHa is quickly removed. Both the single exponential extrapolation and the energy variance (inset) indicate convergence.  
} 
\label{Fig.FeO_Variance} 
\end{figure}

Our results are shown in Figs.~\ref{Fig.Transition_metal} and Figs.~\ref{Fig.FeO_Variance}; the final total energy, together with the signal-to-noise ratio of the Metropolis release constraint at  $\beta$ of final convergence, are listed in the appendix, with the corresponding ph-AFQMC and exact energies for comparison. For all tests, the Metropolis release constraint calculations reduce the error in ph-AFQMC to less than $1$ mHa, even for single determinant $|\Psi_T\rangle$. For Cu+, Metropolis release constraint with the truncated CASSCF $|\Psi_T\rangle$ (of $\sim 300$ determinants ) converges faster than the one with RHF, although not by a dramatic amount. Given the discrepancy in computational cost, the calculation with the RHF 
$|\Psi_T\rangle$ is effectively more efficient. This trend is not universal, however. In FeO, an RHF $|\Psi_T\rangle$ requires a much longer $\beta$ to converge than the CASSCF $|\Psi_T\rangle$ results. In Fig.~\ref{Fig.FeO_Variance} we show the convergence of FeO with a multi-determinant $|\Psi_T\rangle$. As discussed in Sec.~\ref{sec:variance}, the energy variance can be computed conveniently in this formalism. The inset shows an example, now with multi-determinant $|\Psi_T\rangle$ (versus UHF in the example in O$_2$). In realistic situations where there is no reference energy available, the combined use of the energy variance (which must approach zero at convergence) together with the asymptotic exponential fit can provide a very useful gauge for convergence. 

\section{\label{sec:level5}Conclusion and outlook}

In this paper, we introduced a computational framework that smoothly combines the outcome of importance-sampled BRW with MCMC. We apply this approach to ground-state AFQMC to enable a constraint release to be performed with Metropolis following a phaseless or constrained path AFQMC calculation. The constraint release, of course, re-introduces the sign or phase problem, thus incurring an exponential scaling in contrast to the low-polynomial scaling of constrained AFQMC. However, our approach eliminates the usual technical difficulties for undoing the constraining importance functions (which typically have zero values) and makes it possible to interface the two types of calculations with no numerical instabilities. We use the constrained BRW to create a set of initial paths, which are then seamlessly passed on to a Metropolis sampling. The result is a robust constraint release method that can provide an internal gauge of the bias from the constraint and systematically improve upon its result.

We have illustrated the behavior of the Metropolis release constraint algorithm in a range of small molecules. Our test cases include main group molecules as well as bond breaking and transition metal containing systems. The method is able to systematically restore chemical accuracy even in outlier situations where deliberately poor trial wave functions are employed. 

We have also discussed the computation of observables that do not commute with the Hamiltonian. Under the Metropolis formalism, it is particularly convenient to correct for the constraint bias in back-propagation, which tends to be less accurate than for the energy. This opens many interesting possibilities for studying properties. 

As mentioned, the approach we discussed is by no means limited to ground-state calculations. Finite-temperature AFQMC generalization is straightforward. The approach also does not have to be limited to auxiliary-field space. The same interface should be possible in coordinate space QMC, for example, between diffusion Monte Carlo (DMC) and path-integral approaches. Indeed the approach can be applied in any situation where it might be advantageous to first perform BRW (for example, to minimize ergodicity problems) and then interface the result to Metropolis sampling. 

\section{\label{sec:level6}ACKNOWLEDGEMENT}
The authors thank Yuan-Yao He for the valuable discussion and Brandon Kyle Eskridge for help with the frozen-core approximation in the HEAT calculations. Z.X. acknowledges partial support from the U.S. Department of Energy (DOE) under grant DE-SC0001303. Z.X.~is also grateful for the support and hospitality of the Center for Computational Quantum Physics (CCQ), where this work was performed. Computing was carried out at the computational facilities at Flatiron Institute. The Flatiron Institute is a division of the Simons Foundation.

\bibliography{cite} 

\section{\label{sec:level8}APPENDIX A: Table of numerical data} 

In Table~\ref{Table.Energy_Table}, we provide 
additional details on all the shown in the main text.
(The HEAT systems are identical to those in 
Ref.~\cite{sukurma2023_benchmark}, and are not repeated here.) For each system we specify the trial wave function $|\Psi_T\rangle$ (and initial wave function 
$|\Psi_I\rangle$ if it is different from $|\Psi_T\rangle$), the final computed energy from 
constraint release relative to the reference energy, 
the signal-to-noise ratio (average phase), the total 
reference energy, as well 
as the corresponding ph-AFQMC result using the 
same $|\Psi_T\rangle$. As mentioned, for testing purposes, we deliberately 
chose poor trial wave functions so that the error in the
ph-AFQMC results are particularly severe.
The reference energies are from 
Refs.~\cite{AFQMC_mol-Gau-oririnal-Alsaidi2006, AFQMC_bondBreaking, simons_material_2020} as discussed 
in further detail in the main text.

\begin{table*}
\caption{Additional details on the systems 
discussed in the main text.  The column Ph-AFQMC-MRC shows 
the discrepancy of the final Metropolis release constraint result, relative to the reference energy
total ground-state energy (shown in the last column).
Statistical error bars are in the last digit(s) and are in parentheses. 
The second-to-last column,  Ph-AFQMC, gives
the initial value from which the MRC starts.
All energies are in units of mHa.
}
\label{Table.Energy_Table}
\begin{tabular}{ccccccc|cccccccc}
\hline \hline
System & &Trial/Initial wf& & Ph-AFQMC-MRC & & Signal-to-noise ratio $(S)$ &  Ph-AFQMC  &  Exact&  \\\hline
CH+    & & RHF    &  &      $-0.3(5)$          & &    $0.6442(5)$    &     $-3.2(3)$      & $-38003.71$       &   \\
N2(4.2Bohr)     & & UHF    &  &        $-0.9(5)$        & &    $0.2036(6)$   &   $-9.3(2)$    & $-108970.09$       &   \\
N2(4.2Bohr)     & & CASSCF($6\textup{e}$, $12\textup{o}$) &  &        $-0.5(2)$        & &    $0.524(1)$    &    $2.7(5)$   & $-108970.09$       &   \\
Cu+    & & RHF    &  &             $-0.8(7)$   & &     $0.435(1)$   &      $-18.7(12)$   & $-196967.23$       &   \\
Cu+    & & CASSCF($10\textup{e}$, $19\textup{o}$) &  &       $-0.4(7)$        & &     $0.448(1)$     &      $-13.4(6)$  & $-196967.23$       &   \\
VO     & & UHF/RHF    &  &            $0.4(11)$    & &     $0.183(1)$  &      $6.1(5)$   & $-87085.77$       &   \\
CrO    & & UHF    &  &            $-0.7(10)$    & &       $0.337(1)$   &      $14.2(9)$  & $-102558.37$       &   \\
FeO    & & CASSCF($12\textup{e}$, $9\textup{o}$)/RHF    &  &       $0.7(7)$          & &      $0.563(1)$   &     $6.9(8)$   & $-139435.99$       &   \\
\hline \hline
\end{tabular}
\end{table*}

\end{document}